\documentclass[sn-mathphys,Numbered]{sn-jnl}%

\usepackage{graphicx}%
\usepackage{multirow}%
\usepackage{amsmath,amssymb,amsfonts}%
\usepackage{amsthm}%
\usepackage{mathrsfs}%
\usepackage[title]{appendix}%
\usepackage{xcolor}%
\usepackage{textcomp}%
\usepackage{manyfoot}%
\usepackage{booktabs}%
\usepackage{algorithm}%
\usepackage{algorithmicx}%
\usepackage{algpseudocode}%
\usepackage{braket}
\usepackage{listings}%
\usepackage{soul}
\usepackage{anyfontsize}

\newtheorem{theorem}{Theorem}%

\def\Tr{{\rm Tr}}
\newcommand{\R}{\mathbb{R}}
 \newcommand{\C}{\mathbb{C}}
 \DeclareMathOperator{\poly}{poly}
 
 \DeclareMathOperator{\logdet}{logdet}

\newtheorem{remark}{Remark}%
\newtheorem{thm}{Theorem}
\newtheorem{definition}[thm]{Definition}
\newtheorem*{thm*}{Theorem}
\newtheorem{cor}[thm]{Corollary}

\newtheorem{lem}[thm]{Lemma}
\newtheorem*{lem*}{Lemma}
\newtheorem{claim}[thm]{Claim}

\newcommand{\norm}[1]{\left\lVert#1\right\rVert}

\newcommand{\be}{\begin{equation}}
\newcommand{\ee}{\end{equation}}
\newcommand{\eea}{\end{eqnarray}}
\newcommand{\bes}{\begin{equation*}}
\newcommand{\ees}{\end{equation*}}
\newcommand{\eeas}{\end{eqnarray*}}

\algrenewcommand\algorithmicrequire{\textbf{Input:}}
\algrenewcommand\algorithmicensure{\textbf{Output:}}

\begin{document}

\title[Quantum algorithms for spectral sums]{Quantum algorithms for spectral sums}

\author*[1,2]{\fnm{Alessandro} \sur{Luongo}}
\email{ale@nus.edu.sg}

\author[3]{\fnm{Changpeng} \sur{Shao}}

\affil*[1]{\orgdiv{Centre for Quantum Technologies}, \orgname{NUS}, Singapore}
\affil*[2]{\orgdiv{Inveriant Pte. Ltd.}, Singapore}

\affil[3]{\orgdiv{Academy of Mathematics and Systems Science, Chinese Academy of Sciences},  \orgaddress{\city{Beijing}, \postcode{100190}, \country{China}}}

\abstract{
We propose new quantum algorithms for estimating spectral sums of positive semi-definite (PSD) matrices. The spectral sum of an PSD matrix $A$, for a function $f$, is defined as $ \Tr[f(A)] = \sum_j f(\lambda_j)$, where $\lambda_j$ are the eigenvalues of $A$. 
Typical examples of spectral sums are the von Neumann entropy, the trace of $A^{-1}$, the log-determinant, and the Schatten $p$-norm, where the latter does not require the matrix to be PSD. The current best classical randomized algorithms estimating these quantities have a runtime that is at least linearly in the number of nonzero entries of the matrix and quadratic in the estimation error. Assuming access to a block-encoding of a matrix, our algorithms are sub-linear in the matrix size, and depend at most quadratically on other parameters, like the condition number and the approximation error, and thus can compete with most of the randomized and distributed classical algorithms proposed in the literature, and polynomially improve the runtime of other quantum algorithms proposed for the same problems.
We show how the algorithms and techniques used in this work can be applied to three problems in spectral graph theory: approximating the number of triangles, the effective resistance, and the number of spanning trees within a graph.}


\maketitle

\section{Introduction}
Spectral sums are matrix quantities that are central to many problems in computational sciences. They can be found in machine learning, computational chemistry, biology, statistics, finance, and many other disciplines~\cite{rue2005gaussian,rasmussen2010gaussian,cichoki2014bregman,gutman2001energy,hardt2012simple,nie2012low,bengtsson2017geometry,alter2000singular,golub1997generalized,dashti2011uncertainty}, 
Given the wealth of applications for spectral sums, there is a significant interest in developing efficient algorithms to estimate these quantities~\cite{han2017approximating,gramarandomized,ubaru2017fast,musco2017spectrum,Boutsidis,wu2016estimating,han}. 

A spectral sum is defined as the sum of the eigenvalues of a matrix after a given function is applied to them. 

\begin{definition}[Spectral sum \cite{han2017approximating, ubaru2017fast}]
Let $A\in \mathbb{R}^{n\times n}$ be a
symmetric matrix with $A=UDU^\dag$ its eigenvalue
decomposition, where $D={\rm diag}(\lambda_1,\ldots,\lambda_n)$
is the diagonal matrix of
the eigenvalues. Let $f:\mathbb{R} \rightarrow \mathbb{R}$ be a function.
The spectral sum of $A$ for the function $f$ is defined as:
\be
\mathcal{S}_f(A) := {\rm Tr}[f(A)] = \sum_{j=1}^n f(\lambda_j).
\ee
\end{definition}

We can have an analogous definition using the singular values. As an example, consider the \emph{logarithm} of the determinant, which is perhaps the most famous example of a spectral sum, with many applications in computational sciences~\cite{rue2005gaussian,rasmussen2010gaussian,cichoki2014bregman,kerenidis2020quantumEM,reynolds2009gaussian}. Interestingly, the conventional definition of the determinant through the Laplace expansion lead to an algorithm of $O(n!)$ operations, lacking computational efficiency~\cite{gilbertstrang}. Fortunately, practical applications often require the logarithm of the determinant, rather than the determinant itself, and algorithms tailored for the former have proven to be more advantageous in such scenarios. 

There are two key ingredients in most classical algorithms for spectral sums: a way to compute the approximation of a matrix function and a stochastic trace estimation technique. It is possible to compute a matrix function bypassing the costly diagonalization of the matrix, by using techniques from approximation theory and linear algebra. This step is better explained by an example. Imagine you want to compute $\Tr[\log(A)]$, for some PSD matrix $A$. Set $B=I-A$, and observe that for a fixed $m_1 \in \mathbb{N}$:

\begin{equation}\Tr[\log(I-B)] \approx \Tr\left[ - \sum_{k=1}^{m_1}\frac{B^k}{k} \right] = - \sum_{k=1}^{m_1} \frac{\Tr[B^k]}{k}.
\end{equation}

\noindent
This trick bypasses the need for diagonalizing $A$, applying the function to the eigenvalues, and computing the sum, by reducing the problem to (repeated) matrix multiplications and stochastic trace estimation algorithms, for which we have efficient randomized algorithm. Most of these algorithms work in the \emph{matrix-vector product model}, formalized in~\cite{rashtchian2020vector,sun2021querying}. In this model we are given access to an oracle, which returns a vector $Ax$, for a PSD matrix $A$ and a vector $x$. These subroutines are a kind of Monte Carlo algorithm, and the estimate of the trace can be seen as random variables whose expected value is the trace of $A$. For example, consider a random vector $z$ whose entries are Rademacher random variables (i.e. $Pr[+1]=1/2=Pr[-1]$ ). It is known~\cite[Lemma 1]{avron2011randomized} that 

\begin{equation}
  \mathbb{E}[z^TAz] = {\rm Tr}[A].
\end{equation}

\noindent This is called the Hutchinson estimator. Several variants of this estimator exist, like the unit vector estimator, the normalized Rayleigh-quotient, and others~\cite{avron2011randomized}. In the Hutchinson estimator --- which was originally formulated to estimate the trace of the inverse --- it is possible to compute confidence regions for the estimators,
(using either ideas from Monte Carlo techniques, or the Hoeffding inequality~\cite{bai1996bounds, avron2011randomized}). The variance of the estimator for a sample $z$ is: 

\begin{equation}
    {\rm Var}[z^TAz] = 2 \left(\|A\|_F^2 - \sum_{i=0}^{m_2} A_{ii}^2 \right).
\end{equation}

Other recent developments in trace estimation techniques include algorithms based on Krylov subspaces. These methods can be used to approximate the trace of a matrix function for a low-rank approximation of the input matrix~\cite{chen2023krylov}. The algorithm XTrace~\cite{epperly2024xtrace} achieves an efficient estimator by carefully selecting the samples used to estimate the trace, following an ``exchangeability principle''. An implication of this idea, which is leveraged in their work, is that an estimator should be a symmetric function of the samples. Importantly, a recent work ~\cite{meyer2021hutch++} shows how to estimate the trace of a PSD matrix with relative error in only $O(1/\epsilon)$ access to an oracle that gives matrix-vector products (see~\cite{persson2022improved} for improvements). Merging ideas from Hutch++ and techniques such as the Nystr{\"o}m approximation~\cite{li2010making,nakatsukasa2020fast}, it is possible~\cite{saibaba2017randomized,persson2023randomized} to approximate matrix functions under low-rank assumptions. For lower bounds in this model, the interested reader is referred to~\cite{jiang2021optimal,roosta2015improved}, while a recent review of classical algorithms for spectral sums, which focuses on the usage of Krylov subspaces can be found in Ref.~\cite{chen2022randomized}.

In the last 30 years, quantum information has emerged as a new paradigm of computation, which is expected to offer faster-than-classical algorithms. In the last decade, a whole ecosystem of algorithms for quantum linear algebra has emerged (e.g. see~\cite{chakraborty2018power, gilyen2019quantum, low2017optimal, chakraborty2022quantum}). In this work, we show how  --- using a single framework --- we can use quantum computers to estimate different spectral sums.

\subsection{Main results}\label{sec:main results}

We present quantum algorithms to estimate the log-determinant, the von Neumann entropy of a graph, the trace of inverse, and the Schatten $p$-norm, with applications in spectral graph theory. As a corollary, we show how to use some of our techniques and results to estimate the number of triangles in a graph, estimate the effective resistance between nodes of an electric network, and count the spanning trees in a graph.

Our algorithms work in the block-encoding model. In this model, we have access to a unitary $U_A$ (and its inverse) which is a block matrix that encodes a matrix $A$ in its top-left corner. More properly, an $(\alpha, \varepsilon)$-block-encoding of $A$ is a unitary that has a matrix $A'$ in its top-left corner such that $\|A - \alpha A'\| \leq \varepsilon$ (see definition~\ref{def:block-encoding}). As in classical algorithms, we leverage the idea that the trace of a square matrix is equal to the sum of its eigenvalues. Thus, we perform trace estimation of a suitably created matrix which we manipulate with quantum singular value transformation techniques(\cite{chakraborty2018power,gilyen2019quantum, low2019hamiltonian, chakraborty2022quantum, tang2024cs, motlagh2023generalized}). Quantum algorithms for computing the (normalized) trace are well-studied (e.g. in the \emph{one clean qubit} model of quantum computation --- see~\cite{shor2007estimating,Cade} for instance), and they generalize to the gate-based model, and in particular in the block-encoding setting, which we do in appendix~\ref{apx:trace estimation subroutines}. We add to the list of useful quantum subroutines to manipulate block-encodings an extension of~\cite[Lemma 16]{chakraborty2022quantum}, which allows to take the product of $k$ preamplified block-encodings (appendix~\ref{appx:usefulquantum} theorem~\ref{thm:prod of k amplified block-encodings}).

In Section~\ref{sec:dqc1} we discuss the relationship between the algorithms presented in this work and the one clean qubit model of computation. In appendix~\ref{apx:trace estimation subroutines} we rewrite and formalize --- using the notation of block-encodings --- an algorithm that returns an estimate of the trace with absolute and relative error. The algorithms that work with relative error use the algorithm that returns an estimate with absolute error, without knowing a lower bound of the quantity being estimated (see~\cite{chowdhury2021computing,subramanian2021quantum}). In our work, we also use subroutines to estimate the trace of $A^TA$ with relative error~\cite{van2020quantum}. We also discuss our work in relation to recent research in \emph{limited-depth quantum computation}: a setting where we want to reap some of the benefits of quantum computing but due to decoherence, we cannot run circuits for the depth that is necessary to execute the full algorithm.

In the following, let $\kappa$ be the condition number of the matrix $A$, and $\alpha$ be a quantity upper bounded by the Frobenius norm of the matrix (both are precisely defined in Section~\ref{section:notation}). The notation $\widetilde{O}(\cdot)$ hides polylogarithmic factors from the asymptotic complexity. For all the algorithms in  Section~\ref{sec:quantumalgorithmsspectral}, we show exact bounds on the maximum tolerable error of the block-encoding that we receive as input for our algorithms. Specifically, for an $(\alpha, \varepsilon)$-block-encoding, we determine the maximum $\varepsilon$ allowable to ensure that our algorithm produces an estimate of the desired quantity with an error of at most $\epsilon$.

While in the manuscript we report all the polylogarithmic factors in the runtime, our results can be summarised and simplified as follows. Recall that an $\epsilon$-absolute (or additive) error on a quantity $a$ is a quantity $\overline{a}$ such that $|a-\overline{a}|\leq \epsilon$ and an $\epsilon$-relative error is defined as $|a-\overline{a}|\leq \epsilon a$. In appendix~\ref{apx:qaccessclassical} we recall how to build block-encodings within the \emph{quantum memory device} model. 
Lastly, while it is not necessary to use a quantum arithmetic model (appendix~\ref{apx:arithmetic model}), doing so simplifies the analysis of certain steps in our work.

\paragraph{Quantum algorithm for log-determinants} 
The importance of algorithms for the (logarithm of the) determinant of an PSD matrix $A$ cannot be stressed enough. The log-determinant is defined as $\logdet(A)= \sum_{j=1}^n \log \lambda_j$. Notable examples can be found in machine learning and other computational sciences~\cite {rue2005gaussian,rasmussen2010gaussian,cichoki2014bregman,kerenidis2020quantumEM,reynolds2009gaussian}. Our result reads as follows.

\begin{thm*}[Informal - Algorithm for estimating log-determinants]
  \label{thm:logdet-svt-main} 
Let $U_A$ be an $\alpha$-block-encoding of an PSD matrix $A \in \mathbb{R}^{n \times n}$.
  There is a quantum algorithm that returns an estimate  $\overline{\logdet(A)}$ 
  which is with $\epsilon$-relative error of $\logdet(A)$ with high probability 
 using $\widetilde{O}\left(\frac{\alpha \kappa }{\epsilon}\right)$ queries to $U_A$.
\end{thm*}

There are many classical algorithms for this problem~\cite{barry1999monte,pace2004,Boutsidis,avron2011randomized,saibaba2017randomized,aune,han,zhang}. We give a quadratic speedup (in the condition number) with respect to the previous quantum algorithm for the log-determinant~\cite{zhao2019quantum,zhao2019compiling}.

\paragraph{Quantum algorithms for Schatten \texorpdfstring{$p$}{p}-norm}

Let $A\in\mathbb{R}^{m \times n}$ be a matrix with singular values $\{\sigma_i\}_{i=1}^{\min(m,n)}$. The Schatten $p$-norm of $A$, for $p \in \mathbb{N}$ is defined as
$\|A\|_p =  \left( \sum_{i=1}^{\min(m,n)} \sigma_i^p  \right)^{1/p}.$ There are many applications for the algorithms that compute the Schatten $p$-norms, for example in  
completion algorithms, theoretical chemistry, image processing, and many others~\cite{netrapalli2014non, gutman2001energy,lu2015individualized,nie2012low,majumdar2011algorithm,xie2016weighted}. We propose two different results for estimating the Schatten $p$-norm of a matrix. The first one reads as follows.

   \begin{thm*}[Informal -  Algorithm for Schatten $p$-norms]
  \label{thm:schatten-main} 
Let $U_A$ be an $\alpha$-block-encoding of a  matrix $A \in \mathbb{R}^{m \times n}$ with $n\leq m$. 
  There is a quantum algorithm that returns an estimate  $\overline{\|A\|_p}$ which is with $\epsilon$-relative error of $\|A\|_p$ with high probability using $U_A$ for 
$\widetilde{O}\left(\frac{p\sqrt{n}\left(\sqrt{2}\|A\|\right)^{p/2} }{\epsilon \|A\|_p^{p/2}} \left(\frac{\alpha}{\|A\|} \right)\right)$
    times if $p$ is even, or $\widetilde{O}\left(\frac{\sqrt{n}\alpha^{1.5}\left(\sqrt{2}\right)^{p/2}\left(\|A\|\right)^{p/2-1}}{\epsilon \|A\|_p^{p/2}}\left(\frac{p}{2} + \kappa\right)\right)$ times if $p$ is odd.
  \end{thm*}

For large values of $p$, we use an efficient polynomial approximation of monomials in $[-1,1]$ that allows to have a quadratic improvement in $\sqrt{2}^p$. This is better formulated in theorem~\ref{thm:Schatten-p norm big p}. We also report in appendix~\ref{apx:numerical schatten} the results of numerical experiments showing the asymptotic scaling of certain parameters on real-world datasets. These numerical experiments show that in some instances, and certain regimes of parameters, of real-world datasets, the parameter that governs the runtime of the quantum algorithm scales more favorably than the worst-case analysis. Our results improve polynomially over previous quantum algorithms for Schatten $p$-norm~\cite{montanaro2015quantum}.

\paragraph{Quantum algorithm for von Neumann entropy of graphs}

We study an algorithm to estimate the von Neumann entropy of a density matrix. Contrary to many other quantum algorithms for this problem, we do not require access to a purification that allows the creation of a block-encoding of the density matrix. This is especially relevant for problem where the density matrix is obtained from graphs, like the Laplacian of a graph. For a graph $G=(V,E)$ a graph Laplacian $\mathcal{L}$ is a PSD matrix defined as  $\mathcal{L} = \Delta(G)-A(G)$. We can associate to $G$ a density matrix $\rho_G = \mathcal{L}/\Tr[\mathcal{L}]$. Then, the von Neumann entropy of the graph $G$ is defined as $H(G) := H(\rho_G) = -{\rm Tr}[\rho_G \log \rho_G]$. In the following $s(G)$ is the number of edges in the graph $G$, i.e. $|E|$. The von Neumann entropy is used in financial data analysis, genomics, complex network analysis, and pattern recognition~\cite{caraiani2014predictive,banerjee2014feature,alter2000singular,Minello,han2012graph,passerini2008neumann}.

\begin{thm*}
[Informal - Algorithm for estimating von Neumann entropy]
Let $U_A$ be an $\alpha$-block-encoding of an PSD matrix $A \in \mathbb{R}^{n \times n}$.
  There is a quantum algorithm that returns an estimate  $\overline{H(A)}$ which is an estimate of $H(A)$ with  $\epsilon$-absolute with high probability using  $\widetilde{O}\left(\frac{n\alpha\kappa }{\epsilon s(G)}\right)$ queries to $U_A$.

  \end{thm*}

There is a vast literature of quantum algorithms for von Neumann entropies of density matrices $\rho$, but they work in a model where one has access to a purification of the density matrix, which results in a $1$-block encoding of $\rho$~\cite{subramanian2019quantum, li2018quantum, distributional,wang2022new,gur2021sublinear}.

\paragraph{Quantum algorithms for trace of the inverse}
For a square matrix $A$, the trace of the inverse is 
$
I(A) = {\rm Tr}[A^{-1}] = \sum_{i=1}^n \lambda_i^{-1}.  
$
The trace of the inverse finds application
lattice quantum chromodynamics,
 generalized cross validation, and uncertainty quantification~\cite{dashti2011uncertainty,golub1997generalized,stathopoulos2013hierarchical}.

  \begin{thm*}[Informal -  Algorithm for trace of inverse]
  \label{thm:logdet-inverse-main} 
  Let $U_A$ be an $\alpha$-block-encoding of an PSD matrix $A\in\mathbb{R}^{n\times n}$.
There is a quantum algorithm that returns an estimate  $\overline{I(A)}$ which is an estimate of $I(A)$ with $\epsilon$-relative error with high probability using   $O\left(\frac{\alpha^2\kappa^2}{\epsilon}\right)$  queries to $U_A$.
  \end{thm*}

To our knowledge, there are only previous works for the trace of the inverse for classical algorithms~\cite{han2017approximating,ubaru2017fast,wu2016estimating}, while there are no previous quantum algorithms for this problem.

\paragraph{Applications}
We discuss three applications of our algorithms in spectral graph theory. First, we show how to use the techniques from the algorithm for computing the Schatten $p$-norms for counting the number of triangles in a graph. We discuss \emph{two} algorithms. Denoting $\Delta(G)$ the number of triangles in $G$, the runtime of the first algorithm is $O(\frac{\alpha n}{\epsilon \Delta(G)})$, and the runtime of the second algorithm is $O(\frac{\alpha \sqrt{n} \kappa}{\epsilon\sqrt{\Delta(G)}})$. Algorithms for counting triangles find many applications in network analysis~\cite{suri2011counting,easley2012networks}.

Second, we show how to give a relative error estimate of the number of spanning trees in a graph $G$, assuming block-encoding access to the graph Laplacian. Our quantum algorithm makes $\widetilde{O}(n \alpha \kappa /\epsilon)$ calls to the block-encoding of the Laplacian matrix of $A$.
There are many applications in machine learning~\cite{meila2000learning}, genomics, and network theory~\cite{wang2014clustering,kirby2016kirchhoff}.

Third, we show how to use the algorithm for estimating the log-determinant for computing the effective resistance between two nodes in a graph using the block-encoding of (a modified) graph Laplacian $\mathcal{L}$ for $\widetilde{O}(n \alpha \kappa /\epsilon)$ times. 
Applications include graph clustering, graph sparsification and analysis, graph neural networks, and many others~\cite{alev2017graph,fortunato2010community,ahmad2021skeleton,katz1953new,spielman2008graph}.

\subsection{Preliminaries and notation}
\label{section:notation}

We assume a basic understanding of quantum computing, and we recommend ~\cite{nielsen2002quantum} for a comprehensive introduction to the subject. 
For a matrix $M$ we denote its transpose-conjugate as $M^\dagger$. For a matrix $A=(a_{ij})_{n\times n} \in \mathbb{R}^{n \times n}$ we write $A=U\Sigma V^\dag = \sum_{i=1}^n \sigma_i u_i v_i^\dag$ as its singular value decomposition,
where $\sigma_i$ are the singular values and $u_i, v_i$ its left and right singular vectors respectively. 
We use $\lambda_i$ to denote the eigenvalues of a matrix. 
We assume that the singular values (and eigenvalues) are sorted such that $\sigma_1$ is the biggest and $\sigma_n$ is the smallest. We recall the following definition of PSD matrix. 

\begin{definition}[Positive semi-definite (PSD)]
    A Hermitian square matrix $A \in \mathbb{C}^{n \times n}$ is said to be positive semi-definite (PSD) i.e. $A \geq 0$ if $x^\dagger A x \geq 0$ for all $x \in \mathbb{C}^n$.  
\end{definition}

In this paper, we use different matrix norms. With $\|A\|_0$ we denote the number of non-zero elements of the matrix $A$, with $\|A\| = \sigma_1$ the biggest singular value of $A$, and with $\|A\|_F = \sqrt{\sum_{i,j=1}^n  |a_{ij}|^2} = \sqrt{\sum_{i=1}^n  \sigma_i^2}$ its Frobenius norm. 
With $s_r$ ($s_c$) we denote the row(column)-sparsity, that is, the maximum number of non-zero entries of the rows (columns). The sparsity $s$ of a matrix $A$ is defined as the number of non-zero entries (i.e. $s:=\|A\|_0$). If the matrix is the adjacency matrix of a graph $G$, we write $s(G)$. For a matrix $L \in \mathbb{R}^{n \times n}$ and $A \subset [n]$ we denote with $L(A)$ the matrix obtained from $L$ by removing the columns and rows indexed by $A$.

With $\kappa(A)$ we denote the condition number of $A$, that is, the ratio between the biggest and the smallest non-zero singular values. When it is clear from the context, we will simply denote $\kappa(A)$ as $\kappa$.

In this work we use the quantum singular value transformation, a technique comprehensively studied in~\cite{chakraborty2018power,gilyen2019quantum, low2019hamiltonian, chakraborty2022quantum, tang2024cs, motlagh2023generalized}. In appendix~\ref{appx:usefulquantum} we report some results on the product of two preamplified block-encodings, which we generalize to the product of $k$ preamplified block-encodings. In that section one can find precise statements for amplitude amplification and other quantum subroutines are also reported there.

In appendix~\ref{appx:polyapprox}, we report some polynomial approximations of useful functions. Note that the theorem that we are using for singular value transformation (e.g. theorem~\ref{thm:arbParity}) requires classical computation for obtaining a description of a quantum circuit. The runtime for computing this description depends on the degree of the polynomial and the precision used to compute certain angles. As this procedure is very fast on a classical desktop computer, resulting in $0$ or negligible error, we will consider it only in the proof of theorem~\ref{algo:logdet-svt}. The interested reader is referred to~\cite{chao2020finding,dong2021efficient} for more information.

\section{Quantum algorithms for spectral sums}\label{sec:quantumalgorithmsspectral}

In the quantum setting, we compute a spectral sum of a matrix $A$ using a methodology analogous to the classical case, but with different components. We work in a \emph{quantum oracle model}, where along with access to a quantum computer, we have also access to a unitary that gives access to the input of the problem. In our case, this unitary is a block-encoding of the matrix $A$. 

\begin{definition}[Block-encoding~\cite{gilyen2019quantum,low2019hamiltonian}]\label{def:block-encoding}
Suppose that $A \in \mathbb{R}^{2^m \times 2^m}$, $\alpha,\epsilon\in \mathbb{R}^+$ and $q\in\mathbb{N}$. Then we say that the $(m+q)$-qubit unitary $U_A$ is an $(\alpha,q,\epsilon)$-encoding of $A$, if:
\begin{equation}
\|A-\alpha (\langle 0|^{\otimes q}\otimes I ) U_A (|0\rangle^{\otimes q}\otimes I)\|
\leq \epsilon.
\end{equation}
\end{definition}

In this setting, we measure the asymptotic complexity of our algorithms in calls to the unitaries of the block-encoding. Creating block-encodings is quite straightforward in the \emph{QRAM model} (i.e. when we have access to a quantum random access memory via a $\mathsf{QMD}$: a quantum memory device~\cite{allcock2023constant}) or in the \emph{sparse query model}. We discuss both cases in appendix~\ref{apx:qaccessclassical}, where we show how to work with non-square matrices. In those setting, it is possible to build an $(\alpha, a, 0)$-encoding of a matrix, for small $a$, and where $\alpha$ can be the Frobenius norm or another function, which we report in Definition~\ref{def:mu} of the appendix.
To build a block-encoding in both models, we have to perform a classical preprocessing that requires linear time (with some polylogarithmic overhead) in the size of the matrices. This time is needed to build a circuit for sparse access, or a data structure that will be stored in the quantum memory.
Build a block-encoding requires $O(\log m)$ queries to a quantum memory device. The depth of those circuits is usually of $O(\log m)$~\cite{jaques2023qram}. Using a quantum computer with access to $\mathsf{QMD}$ the runtime of the computation is obtained from the depth of the circuits and the number of calls to the $\mathsf{QMD}$. Hence, we use the query complexity of the block-encoded matrix (i.e. number of usages of $U_A$) as a proxy for the runtime of the computation. A precise definition of a quantum computer with access to quantum memory can be found in~\cite{allcock2023constant}.

It will be convenient to assume quantum access to sub-normalized matrices, i.e. where $\|A\|<1$,  say $1/e$. Again, this assumption can be easily satisfied by dividing the matrix by a multiple of its biggest singular value, before creating quantum access to the matrix. For instance, one can use \cite[Algorithm 4.3, Proposition 4.8]{kerenidis2020quantum} to estimate $\|A\|$ with relative error, or $\kappa(A)$ with additive error in time $O(\frac{\|A\|_F\log(1/\epsilon)}{\|A\|\epsilon})$. We assume that this procedure is done immediately after having built quantum access (which, we recall, requires $\widetilde{O}(\|A\|_0)$ time and quantum-accessible space) so the cost of this operation can be subsumed into the cost of creating quantum access to $A$.

In the quantum algorithm we compute a polynomial approximation of the function of choice, and perform quantum singular value transformation on the block-encoding, (for example, using theorem~\ref{thm:arbParity}). In particular, we need to build a block-encoding of $f(A)$ by using singular value transformation techniques~\cite{chakraborty2018power,gilyen2019quantum, low2019hamiltonian, chakraborty2022quantum, tang2024cs, motlagh2023generalized}. For example, we can use the following theorem. 

\begin{thm}[{SVT of Hermitian matrices \cite[Theorem 31]{gilyen2019quantum}}]
\label{thm:arbParity}
		Suppose that $U$ is an $(\alpha,q,\epsilon)$-encoding of a Hermitian matrix $A$. 
		If $\nu\geq 0$ and $ P_{\Re}\in\R[x]$ is a degree-$d$ polynomial satisfying that
		for all $x\in[-1,1]\colon$ $| P_{\Re}(x)|\leq 1/2$.
		Then there is a quantum circuit $\tilde{U}$, which is an $(1,q+2,4d\sqrt{\epsilon/\alpha}+\nu)$-encoding of $ P_{\Re}(A/\alpha)$, and consists of $d$ applications of $U$ and $U^\dagger$ gates, a single application of controlled-$U$ and $O((q+1)d)$ other one- and two-qubit gates.
		Moreover we can compute a description of such a circuit with a classical computer in time $O(\poly{d,\log(1/\nu)})$.
	\end{thm}
In the appendix we report some polynomial approximation for the functions we need: $1/x$, $\log(x)$, and  $x^p$ (i.e. lemma~\ref{lemma:polynomial of inverse}, lemma~\ref{lemma:poly approx ln distributional}, lemma~\ref{lemma:poly approx of monomial}). Other algorithms for manipulating the spectrum of a matrix are reported in appendix~\ref{appx:usefulquantum}. Lastly, we use a quantum algorithm to estimate the trace of a block-encoded matrix.

\begin{lem}[Quantum trace estimation]
Let $\epsilon \in (0,1)$. 
Let $U$ be an $(\alpha, q, \delta)$ block-encoding of $A \in \mathbb{C}^{n \times n}$. There is a quantum algorithm that returns $\overline{{\rm Tr}[A]}$ with  probability at least $2/3$ such that $\left|{\rm Tr}[A] - \overline{{\rm Tr}[A]}\right| \leq  n \epsilon$  using $O\left(\frac{\alpha}{\epsilon}\right)$ calls of $U$ and $U^\dagger$, if $\delta \leq \epsilon/2$.
The probability can be made bigger than $1-\varepsilon$ for $\varepsilon \in (0, 1/2)$ with a multiplicative factor of $O\left(\log(1/\varepsilon) \right)$.
\end{lem}

The proof of this statement can be found in lemma~\ref{lemma:quantum trace estimator vanilla} of appendix~\ref{apx:trace estimation subroutines}, along with an algorithm for estimating a trace with relative error, which we formalize in the language of block-encodings from~\cite{chowdhury2021computing,subramanian2019quantum}. There are other quantum algorithms related to the problem of trace estimation. For example, in~\cite{quek2024multivariate} they estimate the trace of the product of $m$ different density matrices (i.e. $Tr[\rho_1, \dots \rho_m]$), in a computational model where all the $\rho_i$ matrices are available at the beginning of the computation. The quantum circuit has constant depth, but at the expense of running the circuit for $O(\frac{1}{\epsilon^2})$ times. We also report a recent algorithm for multivariate trace estimation~\cite{yosef2024multivariate}, where they also propose high-level set of subroutines to write new quantum algorithms for linear algebra (quantum Matrix State Linear Algebra).

We recall in appendix~\ref{apx:arithmetic model} the \emph{quantum arithmetic model} introduced in~\cite{alphatron,doriguello2022quantum}, which specifies a standard for representing fixed-point numbers. In~\cite{subramanian2019quantum}, they revisited and simplified the previous algorithm by demonstrating how to keep the failure probability of every iteration constant, (but as a function of a lower bound of the trace). The approach of maintaining a constant failure probability for every iteration works well with the quantum arithmetic model, as there exists an inherent ``smallest trace'' that can be estimated on a quantum computer when the numbers are represented in an explicit arithmetic model. We discuss this in lemma~\ref{lem:quantum trace estimator relative} in the appendix.

 In the two sections, we briefly recall two possible venues of study for executing these circuits on real hardware on medium-term quantum computers. The algorithms that we present in this work share a similar structure: the circuit consist in performing amplitude estimation (See~\cite{brassard2002quantum} or theorem~\ref{thm:amplitude_estimation} in the appendix) of a circuit called the Hadamard test, for which we detail the proof in appendix~\ref{apx:trace estimation subroutines}. First, we discuss similarities and differences between the one clean qubit model of computation and the algorithms discussed in this work. Second, we conclude with a concise review of research in \emph{limited-depth quantum computation}: a setting where want to reap benefits of amplitude estimation but due to decoherence we cannot run circuits for the depth that is necessary to execute the full algorithm. In this setting, it is possible to have quantum-classical trade-offs that retain some quantum advantage over the classical algorithms.

\subsection{The one clean qubit model of computation}\label{sec:dqc1}
The DQC1 --- \emph{deterministic quantum computation with one clean qubit} --- is the complexity class of decision problems that can be solved in polynomial time with error probability which scales at most inverse polynomially in the input size on ``one clean qubit'' machines. This is a computational model where we have access to a computer with a single clean (pure) qubit along with a set of qubits in the maximally mixed state. This model of computation was originally motivated by works in NMR (nuclear magnetic resonance) quantum computing~\cite{knill1998power}. A famous complete problem for this complexity class is deciding if the normalized trace of a unitary matrix is greater than a given threshold. It is simple to see that this problem is contained in DQC1~\cite{shor2007estimating}, as it is solved by running a circuit called Hadamard test on a one clean qubit machine. Predictably, it is possible to reduce the trace estimation problem to the decision problem (using a simple binary search).
To see that the problem is DQC1-hard, we can show that estimating the normalized trace is a particular case of the DQC1-complete problem of estimating the coefficient of the Pauli decomposition of a quantum circuit up to polynomial accuracy~\cite{knill1998power,shepherd2006computation,shor2007estimating}. Remarkably, according to commonly accepted assumptions in computational complexity, solving DQC1-complete problems on a classical computer is believed to be hard
\cite{jordan2008quantum,morimae2014hardness,fujii2018impossibility}. However, it is also known that this class is less powerful than a full-fledged quantum computer~
\cite{ambainis2000computing}, and that one clean qubit computer without entanglement can be classically simulated~\cite{yoganathan2019one}. 
We also recall that the class of DQC1 problems is unaffected by extending the one clean qubit model by having access to a logarithmic number of clean qubits~\cite{shepherd2006computation}. We note that one clean qubit quantum computer exhibits noise resilience properties~\cite{goktacs2020benchmarking,datta2007role, datta2008quantum, wang2019witnessing}, which could potentially be leveraged to execute these classes of algorithms on medium-term quantum computers. For the sake of completeness, we recall that it is always possible to recreate a one clean qubit quantum computer on circuit-based architectures. In fact, it is simple to recreate the initial state of a one clean qubit quantum computer using $2n+1$ qubits of which $2n$ are used to create a maximally mixed state  (i.e. by creating a maximally entangled state between two registers of $n$ qubits, and discarding one of the two registers) and using the remaining qubit as the ``clean'' qubit. While it is not clear if this can lead to any computational advantage, this has already been shown experimentally in~\cite{karimi2023power}. 
The interested reader is directed to consult references
\cite{passante2012experimental,xin2018nuclear,jones2024controlling} for further information and recent results on DQC1 and one clean qubit machines.

\subsection{Low-depth algorithms for amplitude estimation}
There is another interesting research direction which has the potential to make the algorithms presented in this work executable on medium-term quantum computers, in particular quantum computers with sufficient number of qubits, but limited depth. In the past, it was shown how to perform a QFT-free amplitude estimation algorithms. While the QFT is not a significant portion of the overall circuit, these techniques eliminate the necessity of using controlled version of the oracle, which is one of the biggest overhead in the execution of the quantum algorithm~\cite{aaronson2020quantum,suzuki2020amplitude,grinko2021iterative}. Among the QFT-free amplitude estimation algorithms, we mention the iterative quantum amplitude estimation~\cite{grinko2021iterative}: perhaps the state-of-the-art result with a provable asymptotic analysis along with a small constant in the asymptotic complexity. For a review and other techniques, the interested reader is encouraged to see~\cite{maronese2023quantum,suzuki2020amplitude}.

Additional advancement has been made in trade-offs between quantum depth and circuit repetitions. In particular, it is possible to interpolate between classical repetitions of a circuit that uses a circuit depth that uses the oracle for a depth that is asymptotically less than $O(1/\epsilon)$, but is repeating the whole quantum circuit more times. The authors in~\cite{giurgica2022low} show how to interpolate between quantum and classical amplitude estimation with an optimal trade-off of 
$ND = O(1/\epsilon^2)$ where $N$ is the number of oracle calls (in our case, the unitary of the block-encoding) and $D$ is the maximum number of sequential oracle calls allowed by our maximum depth. Using these techniques, the algorithms presented here can go from a depth of $O\left(d\frac{1}{\epsilon} \right)$ (where $d$ is the depth of the oracle we are applying amplitude estimation on) to a depth of  $O\left(d\frac{1}{\epsilon^{1-\beta}} \right)$, but running the quantum circuits for $O\left(\frac{1}{\epsilon^{1+\beta}} \right)$ times.

\subsection{Log-determinant}
As mentioned before, the log-determinant of a PSD matrix is the spectral sum where the function we apply to the spectrum is $f(x)=\log(x)$.

\begin{definition}[Log-determinant of an PSD matrix $A$] Let $A \in \mathbb{R}^{n\times n}$ be a PSD matrix, and $b\in \mathbb{N}$. Let $\lambda_1,\ldots,\lambda_n$ be the eigenvalues of $A$. Then the log-determinant of $A$ is defined by
\begin{equation}
\logdet(A) := \log_2 \det (A) = \sum_{j=1}^n \log \lambda_j.
\end{equation}
\end{definition}

An algorithm for estimating the log-determinant can be used to estimate the determinant (and vice versa).
Observe that, while the determinant of a PSD matrix is always positive (because the eigenvalues are all positive), the log-determinant of a non-PSD matrix can be either positive or negative. Under the assumption that the singular values of the matrix lie in the interval $(0,1]$, the log-determinant is always a negative quantity. In case they are not, we can always consider a rescaled matrix $A'=A/\beta$ where  $\beta \geq \norm{A}$. Then, we can recover the log-determinant of $A$ as $\logdet(A) = n \log (\beta) + \logdet(A')$. 

\subsubsection{Applications and previous works} The log-determinant is used in many different research areas. The log-determinant is used for the calculation of the marginal log-likelihood of non-parametric kernel-based methods\cite{dong2017scalable}. For example, computing a log-determinant is necessary when training 
Gaussian processes and Gaussian graphical models~\cite{rue2005gaussian,rasmussen2010gaussian}, and in tasks such as model selection, and model inference. The log-determinant also appears in other machine learning problems, as the computation of Bregman divergences~\cite{cichoki2014bregman}, and quantum and classical algorithms for fitting Gaussian mixtures with Expectation-Maximization algorithms~\cite{kerenidis2020quantumEM,reynolds2009gaussian}.

There are many popular algorithms for the estimation of the log-determinant. For many years, most of the software used algorithms based on the Cholesky decomposition of a matrix. One of the first stochastic algorithms proposed the idea of using a Taylor expansion of the logarithm, along with a stochastic trace estimator subroutine~\cite{barry1999monte}. Few years later, a new algorithm used the Chebyshev approximation of the logarithm function and an exact trace calculation algorithm~\cite{pace2004}. More recently~\cite{Boutsidis}, improved the estimation of the log-determinant using again a Taylor expansion of the logarithm function, but using some improved results~\cite{avron2011randomized} for stochastic trace estimation. Under low-rank assumptions, it is possible to study algorithms based on subspace-iteration and access the matrix only through matrix-vector products~\cite{saibaba2017randomized}. Using iterative methods over Krylov subspaces and dynamic choice of the probing vectors it is possible to improve experimentally the performances of the estimator~\cite{aune}. To our knowledge, the best asymptotic complexity for classical algorithms~\cite{han}  is of  $O(\frac{\sqrt{\kappa}\|A\|_0}{\epsilon^2}\log(\frac{\kappa}{\epsilon\delta}))$, where $\epsilon$ is the approximation error and $\delta$ is the failure probability. This work combines stochastic trace estimators and Chebyshev approximation techniques.  Later on, early results featured some error compensation schemes for improving the accuracy~\cite{zhang} and thus reducing the constant factors in the runtime.

A quantum algorithm for estimating the determinant exists~\cite{zhao2019compiling}, whose complexity can be measured in the number of samples to a quantum algorithm, and has a quadratic dependence on the precision of the results. There is a quantum algorithm for estimating the log-determinant~\cite{zhao2019quantum}, with a complexity of $\widetilde{O}(\frac{\kappa^2 \|A\|_0}{\epsilon})$.

\subsubsection{Quantum algorithm}

\begin{thm}[Algorithm for estimating log-determinants]
  \label{thm:logdet-svt} 
  Let $\epsilon \in (0, 1/6),\delta \in(0,1)$, and let $U_A$ be an $(\alpha, q, \epsilon_1)$-encoding of a PSD matrix $A\in\mathbb{R}^{n\times n}$ with $\|A\| < 1/e$  and $\epsilon_1 \leq \frac{\epsilon^2}{\alpha[64\kappa\log(\alpha\kappa)\log(\frac{\log(\alpha\kappa)}{\epsilon})]^2}$. Then, algorithm~\ref{algo:logdet-svt} returns an estimate  $\overline{\logdet(A)}$ such that $|\overline{\logdet(A)}- \ln(\det(A))|\leq \epsilon |\ln(\det(A))|$ with probability at least $1-\delta$, using $O\left(\frac{\alpha \kappa }{\epsilon}\log(\frac{\log(\kappa\alpha)}{\epsilon})\log(1/\delta)\right)$ calls to $U_A$ and $U_A^\dagger$. 
  \end{thm}

 \begin{algorithm}
\caption{Quantum algorithm for log-determinant}
\label{algo:logdet-svt}
\begin{algorithmic}[1]
\Require 
\Statex An $(\alpha, q, \epsilon_1)$-encoding of a matrix $A \in \mathbb{R}^{n \times n}$ with $\|A\|< 1/e$.
\Statex Error and failure probability $\epsilon, \delta \in(0,1)$. 
\Ensure $\overline{\log\det(A)}  \in \mathbb{R}$

\State Using theorem~\ref{thm:arbParity}, prepare a block-encoding of $\tilde{S}(A/\alpha)$, where $\tilde{S}(x)$ is the polynomial approximation in~ lemma~\ref{lemma:poly approx ln distributional} with $\beta = \frac{1}{\kappa\alpha}$ and error $\frac{\epsilon}{4\log(2\kappa\alpha)}$.  

\State Run the trace estimation algorithm of lemma~\ref{lemma:quantum trace estimator vanilla} with absolute error $\frac{\epsilon}{4\log(2\kappa\alpha)}$ and failure probability $\delta$ on the block-encoding, and call this estimate $L''$.

\State \Return  $\overline{\log\det(A)}=2\log(2\kappa\alpha)L''+n\log\alpha$.
\end{algorithmic}
\end{algorithm}

\begin{proof}
Let $\tilde{S}$ be the polynomial approximation of the function $\frac{\log(x)}{2\log(\kappa)}$ obtained from lemma~\ref{lemma:poly approx ln distributional}, for an error $\epsilon_{SVT}\leq\epsilon/4\log(2\kappa\alpha)$ and $\beta = \frac{1}{\kappa\alpha}$. We can use $\tilde{S}$ in theorem~\ref{thm:arbParity}, so to create a $(1, q+2, 4d\sqrt{\epsilon_1/\alpha}+\nu)$-encoding of $\tilde{S}(A/\alpha)$ 
using $d=O(\alpha \kappa \log(1/\epsilon_{SVT}))$ calls to $U_A$ and its inverse, where $d$ is the degree of the polynomial approximation of lemma~\ref{lemma:poly approx ln distributional}). Note that using the hypotheses on $\epsilon_1$, and setting $\nu=\frac{\epsilon}{16\log(2\kappa\alpha)}$ 
we have now a $(1, q+2, \frac{\epsilon}{8\log(2\alpha\kappa)})$-encoding of $\widetilde{S}(A/\alpha)$. The description of the circuit can be obtained in time $O({\rm poly}(d,\log(1/\nu))$. We run the trace estimation algorithm (lemma~\ref{lemma:quantum trace estimator vanilla}) on the block-encoding, setting the failure probability $\delta$ and absolute error $\epsilon/4\log(2\kappa\alpha)$. This returns an estimate $L''$ of $L'=\Tr[\widetilde{S}(A/\alpha)]$ such that 
$\left| L'' - L' \right|  \leq \frac{n\epsilon}{4 \log(2\kappa \alpha)}$. On the other hand, the error on $L = \Tr\left[\frac{\log(A/\alpha)]}{2\log(2\alpha\kappa)}\right]$ introduced by the polynomial approximation can be bounded as:
\be
\left| L'- L\right| \leq \sum_{i=1}^n \left| \frac{\log(\sigma_i/\alpha)}{2\log(2\alpha\kappa)} - \widetilde{S}(\sigma_i/\alpha) \right| \leq n\epsilon_{SVT} = \frac{n\epsilon}{4\log(2\kappa\alpha)}.
\ee

Our estimator is defined as $\overline{\logdet(A)} :=2\log(2\kappa\alpha)L''+n\log\alpha$. This estimator has an error
$\left|\logdet(A) - \overline{\logdet(A)} \right| $ that can be bounded using the triangle inequality as $2\log(2\kappa\alpha)(\left| L' - L \right| +  \left| L'' - L' \right|) $. Substituting the bounds we computed before,
we get that $\left| \overline{\logdet(A)} - \logdet(A) \right| \leq n\epsilon$.  This circuits uses $U_A$ and its conjugate for $O\left(\frac{\alpha\kappa}{\epsilon}\log(\frac{\log(\kappa\alpha)}{\epsilon})\log(\frac{1}{\delta})\right)$ times. Note that under the hypothesis $\|A\|<1/e$,  we have that $\log(\sigma_i) \leq -1$, so we can bound $n \leq |\logdet(A)|$, so the final quantity can be bounded with a relative error.
\end{proof}

 In the case where $\|A\|>1$ we can scale the matrix as described before. With algorithm \ref{algo:logdet-svt},
we can approximate the log-determinant up to precision $n\epsilon$ (by skipping the observation that $n \leq |\logdet(A)|$). In this case, it is not easy to transform it into a relative error because of the mixed-sign of the logarithm of the singular values.

\subsection{Schatten \texorpdfstring{$p$}{p}-norm}
\label{sec:schattenp}
In linear algebra, a very important matrix norm is the Schatten $p$-norm, which corresponds to the spectral sum for $f(x)=|x|^p$.

\begin{definition}[Schatten $p$-norm~\cite{kittaneh1985inequalities}] 
Let $A\in\mathbb{R}^{m \times n}$ be a matrix with singular values $\{\sigma_i\}_{i=1}^{\min(m,n)}$. Let $p \in \mathbb{N}^+$, then the Schatten $p$-norm $\|A\|_p$ is defined as:
\begin{equation}
\|A\|_p :=  \left( \sum_{i=1}^{\min(m,n)} \sigma_i^p  \right)^{1/p}.
\end{equation}
\end{definition}

Observe that this norm can be defined also for non-square matrices. When $p=1$, it is also known as the nuclear or trace norm, and for $p=2$ the Schatten $2$-norm is the Frobenius norm.

\subsubsection{Applications and previous works}
Schatten $p$-norms are often used in matrix completion algorithms~\cite{netrapalli2014non}, in theoretical chemistry~\cite{gutman2001energy}, rank aggregation and collaborative ranking~\cite{lu2015individualized}, convex relaxations for rank-constrained optimization~\cite{nie2012low}, and in image processing~\cite{majumdar2011algorithm,xie2016weighted}.
There are quantum algorithms for estimating the Schatten $p$-norms. A recent paper~\cite{Cade} focuses on the one clean qubit model, and they assume that both $p$ and $1/\epsilon$ are of order $O({\rm poly}\log n)$. Another quantum algorithm for estimating Schatten $p$-norms~\cite{zhao2019compiling} is similar in spirit to the quantum algorithm for the log-determinant of~\cite{zhao2019quantum}, i.e., they sample from the uniform distribution of singular values and perform the rest of the computation classically. Because of the sampling technique employed there, the algorithm has a dependence on the precision of $O(\epsilon^{-3})$, and further depends quadratically on the Lipshitz constant $K_p$ of the function $x^p$ in the domain $[0, \sigma_{\max}]$.
Furthermore, their algorithms assume that the spectrum of the matrix follows a uniform distribution, i.e. that the singular values are all $\Theta(1)$, which is not the case for low-rank matrices that are obtained from real-world datasets~\cite{udell2019big}.  The present work offers a polynomial speedup with respect to those results. To our knowledge, the fastest classical algorithm for estimating the Schatten-$p$ norm~\cite{han2017approximating} has complexity $\widetilde{O}(\|A\|_0p\kappa/\epsilon^2)$. 
Note that a fast quantum algorithm for estimating the Schatten $2$-norm is lemma \ref{corollary: trace estimation of product final}. If the matrix is PSD, we can use lemma~\ref{lemma:quantum trace estimator vanilla} to estimate the Schatten $1$-norm.\\

\subsubsection{Quantum algorithm}
In theorem~\ref{thm:Schatten-p norm} we consider the case for non-square and non-normalized matrices. In theorem~\ref{thm:Schatten-p norm big p} we propose another quantum algorithm with a further quadratic speedup in $p$, where further assume that  $\|A\| \leq 1$.  In appendix~\ref{apx:numerical schatten} we numerically verify the scaling of the factor $\rho=\frac{(\sqrt{2}\|A\|)^{p/2}}{\|A\|_p^{p/2}}$, which appears in the algorithm's runtimes.  To prove the following theorems we extensively use results on composition on block-encoding: in appendix~\ref{appx:usefulquantum} we propose a new theorem for obtaining the product of $k$ different amplified block-encodings.

\begin{thm}
[Algorithm for estimating Schatten $p$-norms]
\label{thm:Schatten-p norm}
Let $\epsilon, \delta \geq 0$, $p \in \mathbb{N}^+$. Assume $U_A$ is an $(\alpha, a, \varepsilon)$-encoding of a matrix $A\in\mathbb{R}^{m \times n}$ with $n \leq m$. There is a quantum algorithm that computes an $\epsilon$-relative approximation of the Schatten $p$-norm of $A$ using $U_A$ and $U_A^\dagger$:
\begin{itemize}
    \item $\widetilde{O}\left(\frac{p\sqrt{n}\left(\sqrt{2}\|A\|\right)^{p/2} }{\epsilon \|A\|_p^{p/2}} \left(\frac{\alpha}{\|A\|} \right)\right)$
    times if $p$ is even, and $\varepsilon \leq \frac{\epsilon \|A\|_p^{p/2}}{4p(\sqrt{2})^{p/2-1}\sqrt{n}\|A\|^{p/2-1}}$,
    \item $\widetilde{O}\left(\frac{\sqrt{n}\alpha^{1.5}\left(\sqrt{2}\right)^{p/2}\left(\|A\|\right)^{p/2-1}}{\epsilon \|A\|_p^{p/2}}\left(\frac{p}{2} + \kappa\right)\right)$ times if $p$ is odd, and $\varepsilon \leq \frac{\epsilon\|A\|_p^{\frac{p-1}{2}}}{8\sqrt{n}\min (2\alpha^{0.5}, (\sqrt{2}\|A\|)^{(p-1)/2})}$.
\end{itemize}
The $\widetilde{O}$ notation hides factor polylogarithmic in $p,\frac{2^{p/2}\|A\|^p}{\|A\|_p^{p/2}}, \sqrt{n}$, $\alpha$, and $1/\epsilon$.
\end{thm}

\begin{proof}
Note that $\|A\|_p^p = \Tr[\left(A^{p/2}\right)^T A^{p/2}]$, so the idea of this proof is to use lemma~\ref{corollary: trace estimation of product final} (trace estimation of $B^TB$) to estimate the trace of a product of matrices. We consider the following two cases.

\emph{Case 1: $p \equiv 0 \mod 2$}. We create a block-encoding of $A^{p/2}$, where the matrix $A^{p/2}$ is defined as the composition of $A^TA\dots$ for $p/2$ times. This matrix is of size $(n \times n)$ if $p/2$ is even, and $m \times n$ if $p/2$ is odd. 
We use theorem~\ref{thm:prod of k amplified block-encodings} (product of $k$ preamplified block-encodings) which is first amplifying $A$, and then combining the preamplified block-encodings of $A$ for $p/2$ times. 
Obtaining a $\left(\left(\sqrt{2}\|A\|\right)^{p/2}, \frac{p}{2}(a+2), \delta_1\right)$-encoding of $A^{p/2}$ has a cost of

\be
T_1=O\left(\frac{p}{2} \left(\frac{\alpha}{\|A\|} \right)\log\left(\frac{p/2(\sqrt{2})^{(p/2)-1}\|A\|^{p/2}}{\delta_1} \right) \right).
\ee
Now we use lemma~\ref{corollary: trace estimation of product final} to get a relative estimate $S$ of $Tr[(A^{p/2})^TA^{p/2}]$ in time $O\left(\frac{\sqrt{n}\left(\sqrt{2}\|A\|\right)^{p/2} }{\epsilon \|A\|_p^{p/2}} T_1\right)$. Note that using the hypotheses on
$\varepsilon$, we can set $\delta_1$ in theorem~\ref{thm:prod of k amplified block-encodings} to $\delta_1\leq \frac{\epsilon \|A\|_p^{p/2}}{4 \sqrt{n}   }$, so we satisfy all the requirements of lemma~\ref{corollary: trace estimation of product final}.

\emph{Case $2$: $p\equiv 1 \mod 2$.}
This case can be handled by combining the preamplified block-encoding of $A^{(p-1)/2}$ with a block-encoding of $A^{1/2}$. As before, for the first part we build a $( (\sqrt{2}\|A\|)^{(p-1)/2}, \frac{p-1}{2}(a+2), \delta_1)$-encoding of $A^{(p-1)/2}$  using $T_1$ applications of $U_A$. For the second term, we apply the theorem for fractional positive powers, theorem~\ref{lemma: be of any matrix power}. This gives us a 
$(2\alpha^{0.5}, a+1, \delta_2$)-block-encoding of $A^{0.5}$ using 
$T_2 = O(\frac{\alpha \kappa}{\|A\|}\log(\frac{\alpha}{\delta_2}))$ applications of $U_A$. The combination of the two block-encoding (theorem~\ref{thm:prod of block-encodings}) gives a $\left(2\alpha^{0.5}\left(\sqrt{2}\|A\|\right)^{(p-1)/2}, [a+1+\frac{p-1}{2}(a+2)], 2\alpha^{0.5}\delta_1 + \left(\sqrt{2}\|A\|\right)^{(p-1)/2}\delta_2 \right)$-encoding of $A^{\frac{p}{2}}$.
To satisfy the requirement of theorem~\ref{corollary: trace estimation of product final} we set $\delta_1 \leq \frac{\epsilon \|A\|_p^{p/2}}{16\sqrt{n}\alpha^{0.5}}$ and 
$\delta_2 \leq \frac{\epsilon\|A\|_p^{p/2}}{8(\sqrt{2}\|A\|)^{(p-1)/2}\sqrt{n}}$. Note that the hypotheses on $\varepsilon$ satisfy the requirements.  This gives a runtime of $O\left(\frac{\sqrt{n}\alpha^{0.5}\left(\sqrt{2}\|A\|\right)^{(p-1)/2}}{\epsilon \|A\|_p^{p/2}}\left(T_1+T_2\right)\right)$ usages of $U_A$, which is

\begin{align}
O\Biggl(\frac{\sqrt{n}\alpha^{0.5}\left(\sqrt{2}\|A\|\right)^{p/2}}{\epsilon \|A\|_p^{p/2}}\frac{\alpha}{\|A\|} \Bigg(p \log\left(\frac{p(\sqrt{2})^{p/2-1}\|A\|^{p/2}\sqrt{n}\alpha^{0.5}}{\epsilon \|A\|_p^{p/2}} \right) \nonumber \\
+\kappa\log\left(\frac{\alpha(\sqrt{2}\|A\|)^{(p-1)/2}\sqrt{n}}{\epsilon\|A\|_p^{p/2}}\right)\Bigg)\Biggr)
\end{align}

For both cases, we bound the error. We have an estimate of $S$ such that
$
\left|S - \norm{A}_p^p \right| 
\leq \epsilon \norm{A}_p^p.
$
Recall that from theorem~\ref{thm:usefulbound} we have that for $a,b\in \mathbb{R}$:  $(a^p-b^p)=(a-b)(\sum_{i=0}^{p-1}a^{i}b^{p-1-i} )$. Hence, we have
\begin{equation}
\left| S^{1/p} - \|A\|_p\right| 
= \frac{|S - \|A\|_p^p|}
{S^{(p-1)/p} + \cdots + \|A\|_p^{p-1}} 
\leq \frac{|S - \|A\|_p^p|}
{\|A\|_p^{p-1}} 
\leq \epsilon \|A\|_p.
\end{equation}

\end{proof}

The previous algorithm is linear in $p$, as the classical algorithm. In the following, thanks to theorem~\ref{lemma:poly approx of monomial} (polynomial approximation of monomials) we can obtain a further quadratic advantage for computing a Schatten $p$-norms. We formulate the algorithm for working with square matrices with $\|A\| \leq 1$, which allows using theorem~\ref{lemma: BE of matrix power hermitian} (positive powers of Hermitian matrices), instead of theorem~\ref{lemma: be of any matrix power} (positive powers of any matrix). This will save a further factor of $\alpha^{0.5}$ in the final runtime. We believe the algorithm can be generalized to non-square matrices (for example using~\cite{gilyen2019quantum}). For simplicity, we assume a $0$-error block-encoding as input. In the following algorithm, we will use $\widetilde{O}(\sqrt{p})$ calls to $U_A$, where the asymptotic notation hides constant and polylogarithmic factors coming from 
theorem~\ref{lemma:poly approx of monomial} and the computation of the description of the circuits in theorem~\ref{thm:arbParity}. For small and even values of $p$ it might be convenient to use the previous algorithm, which is using exactly $p/2$ calls to $U_A$.

\begin{thm}
[Algorithm for estimating Schatten $p$-norms - big $p$]
\label{thm:Schatten-p norm big p}
Let $\epsilon, \delta \geq 0$, $p \in \mathbb{N}^+$. Assume $U_A$ is an $(\alpha, a, 0)$-encoding of a matrix $A\in\mathbb{R}^{n \times n}$ such that $\|A\| \leq 1$. There is a quantum algorithm that computes an $\epsilon$-relative approximation of the Schatten $p$-norm of $A$ using $U_A$ and $U_A^\dagger$:
\begin{itemize}
    \item $\widetilde{O}\left(\frac{\alpha\sqrt{np}}{\epsilon}\frac{(\sqrt{2})^{p/2}\|A\|^{p/2-1}}{\|A\|_p^{p/2}} \right)$ times if $p$ is even,
    \item $\widetilde{O}\left(\frac{\sqrt{n}\left(\sqrt{2}\|A\|\right)^{(p-1)/2}}{\epsilon \|A\|_p^{p/2}} \left(\frac{\alpha\sqrt{p}}{\|A\|} + \alpha\kappa \right) \right)$ times if $p$ is odd.

\end{itemize}
    The $\widetilde{O}$ notation hides factor polylogarithmic in $n,p,\frac{2^{p/2}\|A\|^p}{\|A\|_p^{p/2}}, \sqrt{n}$, and $1/\epsilon$.
\end{thm}

\begin{proof}
Again, as $\|A\|_p^p = \Tr[\left(A^{p/2}\right)^T A^{p/2}]$, we use lemma~\ref{corollary: trace estimation of product final} (trace estimation of $B^TB$) to estimate the trace of a product of matrices. We consider the following two cases.

\emph{Case 1: $p=0 \mod 2$}. For 
$\delta \leq \frac{\sqrt{2}\|A\|\epsilon^2\|A\|_p^p}{256np\log\left(\frac{12\|A\|^pn}{\epsilon\|A\|_p^p}\right)}$ we amplify $U_A$ to a $(\sqrt{2}\|A\|, a+1, \delta)$-encoding, with a costs of $O(\frac{\alpha}{\|A\|}\log(1/\delta_1))$, and we create a $(1, a+3, 4d\sqrt{\delta_1/(\sqrt{2}\|A\|)} + \nu)$-encoding of $\left(\left(\frac{A}{\sqrt{2}\|A\|}\right)^{p/2}\right)/2$ using theorem~\ref{thm:arbParity} (SVT of Hermitian matrices) where we set $\nu
$ of theorem~\ref{thm:arbParity} to be $\nu \leq \frac{\epsilon \|A\|_p^{p/2}}{8 \sqrt{n}}$. To obtain this block-encoding we use the polynomial $E$ of degree $d= \lceil \sqrt{p\log(\frac{12\|A\|^{p/2}n}{\epsilon \|A\|_p^p})} \rceil$ obtained from theorem~\ref{lemma:poly approx of monomial} (efficient approximation of monomials), which can approximate the monomial $\frac{x^{p/2}}{2}$ with error $\epsilon' \leq \frac{\epsilon \|A\|^p_p}{2(3\|A\|^{p/2}n)}$. Note that we have the following relative error in the approximation of the trace:

\begin{align}
\left|\|A\|_p^p - \Tr[E(A)^TE(A)]\right| \leq & \sum_{i=1}^n \left|\sigma_i^{p} - E(\sigma_i)E(\sigma_i) \right| 
\leq  \sum_{i=1}^n \left|\sigma_i^p - (\sigma_i^{p/2}\pm \epsilon' )^2 \right|  \nonumber \\
& \leq \left(2\|A\|^{p/2}\epsilon'+\epsilon'^2 \right)n \leq  \frac{\epsilon \|A\|_p^p}{2},
\end{align}
where we upper bounded $\epsilon'^2$ by $\|A\|^{p/2}\epsilon'$. 
The number of calls to $U_A$ for producing this block-encoding becomes: 

\be
T_1=O\left(\frac{\alpha}{\|A\|}\sqrt{p\log\left(\frac{\|A\|^{p/2}n}{\epsilon \|A\|_p^p}\right)} \log(1/\delta_1) \right).
\ee
It is simple to see that the block-encoding we built is a $(2(\sqrt{2}\|A\|)^{p/2}, a+3, \frac{\epsilon\|A\|_p^{p/2}}{4\sqrt{n}})$-encoding of $A^{p/2}$. Thus, now we can use lemma~\ref{corollary: trace estimation of product final} to get a relative estimate $S$ of $Tr[(A^{p/2})^TA^{p/2}]$ with error $\frac{\epsilon\|A\|_p^p}{2}$ in time

\begin{equation}
    O\left(\frac{\sqrt{n}(\sqrt{2}\|A\|)^{p/2}}{\|A\|_p^{p/2}\epsilon}\frac{\alpha}{\|A\|}\sqrt{p \log \left(\frac{\|A\|^{p/2}n}{\epsilon\|A\|_p^p}\right)}\log\left(\frac{np\log\left(\frac{\|A\|^pn}{\epsilon\|A\|_p^p}\right)}{\|A\|\epsilon^2\|A\|_p^p}\right) \right)
\end{equation}
The error in the final estimate is given by the trace estimation ($\frac{\epsilon\|A\|_p^p}{2}$) and the polynomial approximation error ($\frac{\epsilon\|A\|_p^p}{2}$), and can be bounded by the triangle inequality by $\epsilon\|A\|_p^p$.

\emph{Case 2: $p=1 \mod 2$.}
We create a block-encoding of $A^{(p-1)/2+\frac{1}{2}}$.First, we build a $( (\sqrt{2}\|A\|)^{(p-1)/2}, a+3, 4d\sqrt{\delta_1/\sqrt{2}\|A\|} + \nu)$-encoding of $A^{(p-1)/2}$  using a preamplified block-encoding of $A$. This will require $O(T_1)$ applications of $U_A$. For the second term, we apply the theorem~\ref{lemma: BE of matrix power hermitian} (positive powers of Hermitian matrices). This gives us a 
$(2, a+\log\log(1/\delta_2), \delta_2$)-block-encoding of $A^{0.5}$ using 
$T_2 = O(\alpha \kappa\log^2(\frac{\kappa}{\delta_2}))$ applications of $U_A$\footnote{we consider the number of ancilla qubits of theorem~\ref{lemma: BE of matrix power hermitian} to be $\lceil\log\log(1/\epsilon) \rceil$}. 
Theorem~\ref{thm:prod of block-encodings} (product of block-encodings) gives a block-encoding of $A^{\frac{p}{2}}$ of the kind:

{\small\begin{equation}
\left(2\left(\sqrt{2}\|A\|\right)^{(p-1)/2}, [a+3+\log\log(1/\epsilon)], 2(4d\sqrt{\delta_1/\sqrt{2}\|A\|} + \nu) + \left(\sqrt{2}\|A\|\right)^{(p-1)/2}\delta_2 \right)
\end{equation}}

By setting $\nu \leq \frac{\epsilon \|A\|_p^{p/2}}{24\sqrt{n}}$, 
 $\delta_2 \leq \frac{\epsilon \|A\|_p^{p/2}}{12 \sqrt{n}(\sqrt{2}\|A\|)^{(p-1)/2}}$ and $\delta_1 \leq \frac{\epsilon\|A\|_p^{p/2}}{96\sqrt{n}d}$, where $d$ is defined as before, we have a $\left(2\left(\sqrt{2}\|A\|\right)^{(p-1)/2}, [a+3+\log\log(1/\epsilon)], \frac{\epsilon\|A\|^{p/2}_p}{4\sqrt{n}} \right)$, and thus we can use theorem~\ref{corollary: trace estimation of product final} to get an estimate $S$ of the trace of the product of the block-encoding with error $\frac{\epsilon\|A\|^{p/2}_p}{2\sqrt{n}}$.
The runtime becomes:

\begin{align}
O\Biggl(\frac{\sqrt{n}\left(\sqrt{2}\|A\|\right)^{\frac{p-1}{2}}}{\epsilon \|A\|_p^{p/2}} 
&\Biggl(\frac{\alpha}{\|A\|}\sqrt{p\log\left(\frac{\|A\|^{\frac{p-1}{2}}n}{\epsilon\|A\|_p^p}\right)}\log\left(\frac{\sqrt{n}d}{\epsilon\|A\|_p^{p/2}}\right)+ \nonumber \\
& \alpha\kappa\log^2\left(\frac{\kappa \sqrt{n}(\sqrt{2}\|A\|)^{\frac{p-1}{2}})}{\epsilon\|A\|_p^{p/2}}\right)\Biggr)\Biggr). 
\end{align}

The error analysis is similar to the error analysis of the even case.
For both cases, the trace estimation subroutines returns an estimate an $S$ such that
$
\left| S - \norm{A}_p^p] \right| 
\leq \epsilon \norm{A}_p^p.
$
The conclusion of the proof is now identical to the proof of theorem~\ref{thm:Schatten-p norm}.
\end{proof}

\subsection{von Neumann entropy of graphs}\label{sec:vonneumannentropy}

The spectral sum for the function $f(x)=-x\log x$ (where we take $0 \log 0 = 0$ by convention) can be linked to the von Neumann entropy of a quantum state: the quantum generalization of the classical Shannon entropy. There is a rich literature that studies the notion of quantum entropy of networks and other combinatorial structures, see for example~\cite{braunstein2006laplacian}. For the following definition, it suffices to recall that a density matrix is defined as a PSD matrix with unitary trace. 

\begin{definition}[von Neumann entropy of a density matrix] For a density matrix $\rho$ with eigenvalue decomposition $\rho = \sum_{i=1}^n \lambda_i \ket{\psi_i}\bra{\psi_i}  \in\mathbb{R}^{n \times n}$, the von Neumann entropy of $\rho$ is defined as:
\begin{equation}
H(\rho) := -{\rm Tr}[\rho \log \rho] = - \sum_{i=1}^n \lambda_i \log \lambda_i.
\end{equation}
\end{definition}

We can associate the notion of entropy of a simple, undirected, and connected graph $G=(V,E)$ with $n$ vertices. Let $A(G)$ be the adjacency matrix of the graph (i.e. $A(G)_{ij} = 1$ if $(i,j) \in E$),  and $\Delta(G)$ be the so-called degree matrix of $G$: a diagonal matrix where $\Delta(G)_{ii} = d(i)$, where $d(i)$ is the number of neighbors of the node $i$. The \emph{combinatorial Laplacian matrix} of $G$ is defined as $\mathcal{L}(G) := \Delta(G)-A(G)$. It is possible to see that the matrix is PSD (see the Gershgorin disk theorem~\cite{horn2012matrix}, or observe that $\mathcal{L}s=M^TM$ where $M$ is the incidence matrix of $G$) and that the rank of the matrix is $n-k$, where $k$ is the number of connected components of the graph~\cite{bapat1996laplacian}. To obtain a density matrix from a graph, we define  $\rho_G :=  \frac{\mathcal{L}}{\Tr[\mathcal{L}]} = \frac{\mathcal{L}}{s(G)}$ where we noted that $s(g) = \Tr[\mathcal{L}]$. The entropy $H(G)$ of a graph $G$ is then defined as the entropy of $H(\rho_G)$.

\begin{definition}[von Neumann entropy of a graph] Let $G=(V,E)$ be a graph with Laplacian $\mathcal{L} =  \Delta(G)-A(G)= \sum_{i=0}^{n-1} \nu_i u_iu_i^T$ and associated density matrix $\rho_G = \mathcal{L}/\Tr[\mathcal{L}] = \sum_{i=0}^{n-1} \lambda_i u_iu_i^T$. The von Neumann entropy of the graph $G$ is defined as:
\begin{equation}
H(G) := H(\rho_G) = -{\rm Tr}[\rho_G \log \rho_G] = -\sum_{i=1}^n \lambda_i \log \lambda_i.
\end{equation}
\end{definition}

Observe that the entropy of $G$ can be  rewritten as 
\begin{align}\label{eq:entropyunpacked}
H(G) & = -\sum_i \lambda_i \log (\lambda_i) = -\sum_i \frac{\nu_i}{s(G)} \log\left( \frac{\nu_i}{s(G)} \right) \nonumber \\ 
    &= -s(G)^{-1} \left(\sum_i \nu_i\log(\nu_i) + \sum_i \nu_i \log(s(G)) \right) \nonumber \\
    & = -s(G)^ {-1} \left( \sum_i \nu_i \log(\nu_i) + \log (s(G)) Tr[\mathcal{L}] \right) \nonumber \\
    & = -s(G)^{-1}\sum_i \nu_i \log \nu_i + \log(s(G)).
\end{align}

\subsubsection{Applications and previous works}
The von Neumann entropy of a quantum state quantifies the amount of entanglement contained in a bipartite quantum system~\cite{preskill1998lecture,bengtsson2017geometry}. In applications, the von Neumann entropy is also important in feature selection~\cite{banerjee2014feature}, financial data analysis~\cite{caraiani2014predictive}, and genomic data~\cite{alter2000singular}.  In graph theory, the von Neumann entropy is a spectral measure that has applications in complex network analysis and pattern recognition~\cite{Minello,han2012graph,passerini2008neumann}. Three randomized classical algorithms were given in~\cite{gramarandomized}, where the best classical algorithm is \cite[theorem 2]{gramarandomized} which shows an asymptotic complexity  of
$\widetilde{O}(\|A\|_0\kappa/\epsilon^{2})$ operations.

There is a vast literature on quantum algorithms for estimating von Neumann entropies, and the more general R\'{e}nyi entropies~\cite{subramanian2019quantum, li2018quantum, distributional}. 
In~\cite{distributional}, they show how to estimate with additive error using $\widetilde{O}(\frac{n}{e^{1.5}})$ queries to the oracle giving the purification.

To cite some recent results, in~\cite{wang2022new}
the algorithms assume quantum access to a purification of a density matrix that encodes a sub-normalized operator. In this setting, they can obtain algorithms that do not depend on the condition number of the operator or the size of the system, but only on the rank of the operator, achieving an asymptotic of $\widetilde{O}(r^2/\epsilon^2)$.
Always in the purified query model, it is also possible to estimate the entropy with a relative error using $\widetilde{O}\left( n^{\frac{1}{2} + \frac{1+\eta}{2\gamma^s}}\right)$ queries, where $\eta$ is a lower bound on the entropy, and $\gamma$ is the relative precision~\cite{gur2021sublinear}.

\subsubsection{Quantum algorithm}
Most of the previous works in quantum algorithms for von Neumann entropies assume to have access to a unitary matrix that produces a state that is the purification of a density matrix, as is described in lemma \ref{lemma:purification} in  appendix~\ref{apx:qaccessclassical}. From a purification is possible to obtain a $1$-block encoding of $\rho$. We can relax this requirement, by assuming a generic $\alpha$-block-encoding of a (normalized) combinatorial Laplacian $L:=\mathcal{L}/\|\mathcal{L}\|$. When working using with graphs in the sparse access model (see appendix~\ref{appx:usefulquantum}) we have quantum access to a matrix whose spectral norm is usually greater than $1$. We can always estimate it, and re-create quantum access to the scaled matrix during the preprocessing.

\begin{thm}
[Algorithm for estimating von Neumann entropy]
\label{thm:von Neumann entropy}
Let $G$ be a graph with graph Laplacian $\mathcal{L}$, for which we know $s(G)$.  Let $\epsilon \in (0, 1/2),\delta \in(0,1)$, and let $U_L$ be a an $(\alpha, q, \varepsilon)$-encoding of $L = \mathcal{L}/\|\mathcal{L}\|$ and $\varepsilon \leq \frac{s(G)^2\epsilon^2\alpha}{(196\sqrt{2}n\|\mathcal{L}\|\kappa \alpha \log(2\kappa\alpha)\log(n\log(2\kappa\alpha/\epsilon)))^2 }$. Then, there is an algorithm that  returns  an $\epsilon$-absolute approximation of  $H(G)$  with probability at least $1-\delta$, using $O\left(\frac{n\|\mathcal{L}\|\alpha\kappa }{\epsilon s(G)} \log^ 2\left(\frac{n\|\mathcal{L}\|\log(2\kappa\alpha)}{\epsilon }\right) \log\left(\frac{1}{\delta}\right)\right)$ calls to $U_L$ and $U_L^\dagger$.
\end{thm}

\begin{proof}
The idea behind this algorithm is to obtain a block-encoding of the function $x\log x$ by combining the (preamplified) block-encoding of the function $x$ (i.e. the block-encoding of the matrix itself) with the block-encoding of the polynomial approximating $\log x$. For the latter, we can use again the polynomial $\tilde{S}(x)$ with error $\epsilon_1$ defined in lemma 
\ref{lemma:poly approx ln distributional}, which combined with $x$ can approximates $F(x) = \frac{-x(\log x/\alpha)}{3\log(2\alpha/\beta)}$ on $[\beta, 1]$ (for $\beta = {1}/{\kappa}$). We call this approximating polynomial $P(x) = -x\tilde{S}(x/\alpha) $. Note that for $x \in [\beta, 1]$ we have that $ |\frac{-x(\log x/\alpha)}{3\log(2\alpha/\beta)} - x\widetilde{S}(x) | \leq x\epsilon_1 \leq \epsilon_1$. This polynomial thus approximates $F(x)$ with error $\epsilon_1$. The degree of the polynomial $P(x)$ is thus $d=O(\kappa\alpha\log(1/\epsilon_1))$ with $\epsilon_1 \leq \frac{\epsilon}{6n\log(2\kappa\alpha)}$. We can use theorem~\ref{thm:arbParity} to build a $(1, q+2, 4d\sqrt{\varepsilon/\alpha})$-block-encoding of $\tilde{S}(L/\alpha)$, and we note that from the hypothesis on $\varepsilon$, this is a $(1, q+2, \frac{s(G)\epsilon}{48\sqrt{2}n\|\mathcal{L}\|\log(2\kappa\alpha)})$-block-encoding of $\widetilde{S}(L/\alpha)$. We combine this block-encoding with $U_L$ using 
theorem~\ref{thm:product preamplified be 2022} (product of preamplified block-encodings). This creates a $(2, 2q+2, \frac{s(G)\epsilon}{12n\|\mathcal{L}\|\log(2\kappa\alpha)})$-block-encoding of $P(L)$ using $U_L$ and $U_L^\dagger$ for

$$O\left(\left[\alpha + \alpha\kappa \log(1/\epsilon_1) \right]\log\left(\frac{n\|\mathcal{L}\|\log(2\kappa\alpha)}{s(G)\epsilon}\right) \right).$$

Now define $T$ as 
\begin{align}
T& =\Tr[F(A)] = - \sum_{i} \frac{\nu_i}{\|\mathcal{L}\|} \frac{\log(\nu_i /\|\mathcal{L}\|\alpha)}{3\log(2\kappa \alpha \|\mathcal{L}\|)} \nonumber  \\
& = -(3\|\mathcal{L}\|\log(2\kappa\alpha \|\mathcal{L}\|))^{-1} \left[ \sum_i \nu_i \log \nu_i - \sum_i \nu_i\log(\|\mathcal{L}\|\alpha)  \right] \nonumber \\
& = -(3\|\mathcal{L}\|\log(2\kappa\alpha \|\mathcal{L}\|))^{-1} \left[ \sum_i \nu_i \log \nu_i - Tr[\mathcal{L}] \log(\|\mathcal{L}\|\alpha) \right].
\end{align}
Conveniently, we also define $b :=\sum_i \nu_i \log \nu_i = \Tr[\mathcal{L}]\log(\|\mathcal{L}\|\alpha) - T(3\|\mathcal{L}\|\log(2\kappa\alpha))$. For $T' = \Tr[P(A)]$, we bound the error with $T$ as 
\begin{align}
|T-T'| & \leq \left| \sum_i \frac{\nu_i}{\|\mathcal{L}\|} \left(\frac{\log(\nu_i/\alpha)}{3\log(2\alpha\kappa)} - \widetilde{S}(\frac{\nu_i}{\alpha}) \right) \right| \leq \frac{n\Tr[\mathcal{L}]\epsilon_1}{\|\mathcal{L}\|} \leq \frac{s(G)\epsilon}{6\|\mathcal{L}\|\log(2\kappa\alpha \|\mathcal{L}\|)}.
\end{align}

We get $T''$ as an estimate of $T'$ by using the block-encoding of $P(L)$ in the trace estimation subroutine of lemma~\ref{lemma:quantum trace estimator vanilla} with absolute error $\leq \frac{s(G)\epsilon}{6n\|\mathcal{L}\|\log(2\kappa\alpha)}$ and failure probability $\delta$, so that $|T-T'|\leq \frac{s(G)\epsilon}{6\|\mathcal{L}\|\log(2\kappa\alpha)}$. Thus, as $|T-T''| \leq \frac{s(G)\epsilon}{3\|\mathcal{L}\|\log(2\kappa\alpha) }$ (by triangle inequality), and by defining $b'$ and $b''$ similarly to $b$, we can bound $|b-b''| \leq \epsilon s(G)$. Recalling eq.\ref{eq:entropyunpacked}, it follows that

\begin{equation}
|H(G) - \overline{H(G)}| = s(G)^{-1}|b - b''| \leq \epsilon.
\end{equation}

To conclude, observe that the final runtime is
$$O\left(\frac{n\|\mathcal{L}\|\alpha\kappa }{\epsilon s(G)} \log\left(\frac{n\log(2\kappa\alpha)}{\epsilon}\right)\log\left(\frac{n\|\mathcal{L}\|\log(2\kappa\alpha)}{\epsilon s(G)}\right) \log\left(\frac{1}{\delta}\right)\right).$$

\end{proof}

\subsection{Matrix inverse}

The trace of a matrix inverse is the spectral sum for the function $f(x)=1/x$.

\begin{definition}[Trace of inverses\cite{han2017approximating}] For a matrix $A\in\mathbb{R}^{n \times n}$ with non-zero eigenvalues $\{\lambda_i\}_{i=1}^n$, the trace of the inverse of $A$ is defined as:
\begin{equation}
I(A) := {\rm Tr}[A^{-1}] = \sum_{i=1}^n \lambda_i^{-1}.  
\end{equation}
\end{definition} 
\noindent Note that for any nonzero $\alpha$, we have $I(\alpha A) = I(A)/\alpha$. Thus, if we can approximate the trace of a matrix inverse of a scaled matrix up to relative error $\epsilon$, we can also obtain an $\epsilon$-relative approximation of the original quantity $I(A)$. 
In most applications, the matrix is usually PSD, but we are not using this assumption in our algorithm.

\subsubsection{Applications and previous works}
Computing this quantity has wide applications, especially in 
  the study of lattice quantum chromodynamics~\cite{stathopoulos2013hierarchical},
 generalized cross validation~\cite{golub1997generalized}, and uncertainty quantification~\cite{dashti2011uncertainty}.

To our knowledge, the best known classical algorithm~\cite{han2017approximating} to solve this problem has complexity $\widetilde{O}(\|A\|_0\sqrt{\kappa}/\epsilon^2)$, and is using Chebychev approximation and stochastic trace estimation algorithms. Another classical algorithm~\cite{ubaru2017fast}, which uses stochastic Lanczos quadrature, has a similar runtime.  Recent work~\cite{wu2016estimating} leverages preconditioners and ideas to approximate the diagonal of the inverse of the matrix using the diagonal of some approximate inverse that can be computed inexpensively. Furthermore, it uses dynamical techniques to control the error of the estimation. However, it lacks a theoretical analysis of the runtime of the algorithms. To our knowledge, there are no previous quantum algorithms for this problem.

\subsubsection{Quantum algorithm}

\begin{thm}[Algorithm for estimating the trace of the inverse]
  \label{thm:trace of inverse} 
  Let $\epsilon \in (0, 1/2),\delta \in(0,1)$, and let $U_A$ be a an $(\alpha, q, \epsilon_1)$-encoding of a PSD matrix $A\in\mathbb{R}^{n\times n}$ with $\|A\| \leq 1$  and $\epsilon_1 \leq \frac{1}{\alpha^3}\left(\frac{3\epsilon}{128\kappa^2\log(\frac{\alpha\kappa}{\epsilon})}\right)^2$. Then, there is an algorithm that returns an $\epsilon$-relative approximation of  $I(A)$  with probability at least $1-\delta$, using $O\left(\frac{\mu^2\kappa^2\log(\mu\kappa/\epsilon)}{\epsilon}\log(1/\delta)\right)$ calls to $U_A$ and $U_A^\dagger$. 
  \end{thm}

\begin{proof}
We perform singular value transformation (i.e., theorem \ref{thm:arbParity}) with the polynomial approximation $\widetilde{V}(x)$  (lemma \ref{lemma:polynomial of inverse}) of the inverse function $f(x)={3\delta}/{8x}$ on the interval $[-1,1] \backslash [-\delta,\delta]$ with error $\epsilon_{SVT} \leq \frac{3\epsilon}{16 \kappa \alpha}$.
Here set $\delta=1/\mu\kappa$, so the degree of $\widetilde{V}$ is $O(\mu\kappa\log(1/\epsilon_{SVT})$.
Using the hypothesis on $\epsilon_1$, the singular value transformation
gives us a $(1, 2+q, \frac{3n\epsilon}{32\kappa\alpha})$-encoding of $\widetilde{V}(\frac{A}{\mu})$. We use this block-encoding in the trace estimation subroutine (i.e. lemma \ref{lemma:quantum trace estimator vanilla}) with absolute error $\epsilon \leq \frac{3\epsilon}{16\kappa\alpha}$. In this way we obtain an estimate $T''$ of $T'=\Tr[\widetilde{V}(\frac{A}{\mu})]$ such that $\left| T' - T'' \right| \leq \frac{3n\epsilon}{16\kappa\alpha}.$ We define our estimate of 
$\overline{I(A)} := 8\mu\kappa T''/3$. The polynomial approximation in lemma \ref{lemma:polynomial of inverse} introduces in the estimate of $T$ an error bounded by
\begin{equation}
\left| T - T' \right| 
\leq \sum_{i=1}^n 
\left|  \widetilde{V}(\sigma_i/\mu)  - \frac{3\kappa}{8 \sigma_i/\alpha}\right| 
\leq  \frac{3n\epsilon}{16\kappa\alpha}.
   \end{equation}

The total error in our estimator follows from the triangle inequality and is:
\begin{align}
\left|I(A) - \overline{I(A)} \right| = &  \frac{8}{3\kappa\alpha}\left| T - T'' \right| \leq  \frac{8\mu\kappa}{3}\left(|T-T'|+|T'-T''|\right)  \nonumber \\
\leq  &  \epsilon n \leq \epsilon I(A).
\end{align}

In the last inequality, we used again the hypothesis that $\|A\| \leq 1$, so $I(A) \geq n$.

The algorithm uses $U_A$ and $U_A^\dagger$ for 
$O\left(\frac{\mu^2\kappa^2\log(\mu\kappa/\epsilon)}{\epsilon}\log(1/\delta)\right)$.

\end{proof}

\section{Applications in spectral graph theory}\label{sec:applications}
We discuss three applications to problems in spectral graph theory. For all of them, we assume we are given a unitary that is an $(\alpha, 0)$-block-encoding.

\subsection{Estimating the number of triangles in a graph}\label{sec:triangles}

A simple application of the algorithms outlined before consists of approximating the number of triangles $\Delta(G)$ in an undirected and unweighted graph $G$ on $n$ vertices. The problem of counting the number of triangles in a graph holds both theoretically and practically significance. 
For instance, it is related to the clustering coefficient of a node in social network analysis, see~\cite{suri2011counting, easley2012networks} for a thorough discussion. Let $A$ be the adjacency matrix of $G$. It is known that: 
\be \label{eq:triangleadjacecny}
\Delta(G) = \frac{1}{6}\Tr[A^3] = \frac{1}{6}\sum_{i=1}^n \lambda_i^3.
\ee

For a symmetric matrix $A$, let  $A = \sum_i^n \lambda_i \ket{u_i}\bra{u_i}$ be its eigenvalue decomposition, and $A=\sum_i^n \sigma_i \ket{u_i}\bra{v_i}$ be its singular value decomposition. 
It is simple to verify that for a symmetric $A$, we have that $\Tr[A^3] = \sum_i \lambda_i^3 =  \Tr[U\Lambda^3 U^T] = \Tr[U\Sigma^3 V^T]$. The last equality holds because of the following observation. First recall that $\sigma_i = |\lambda_i|$. Then, by definition of SVD, we have that $\ket{u_i} = \frac{1}{\sigma_i}A\ket{v_i}$, for each $\ket{v_i}$. But as $\ket{v_i}$ is an eigenvector of $A$, we also have that 
$\ket{u_i} = \frac{\lambda_i}{|\lambda_i|}\ket{v_i}$. From this we can rewrite $U\Lambda U^T = U\Sigma S U^T$, where the matrix $S$ is a diagonal matrix having on the diagonal $s_{ii} = \text{sign}(\lambda_i)$. The equality follows from the cyclic property of the trace.

Quantum algorithm for triangle finding exists~\cite{hamoudi2018quantum}. 
Their algorithm works in the general graph model, gives a relative error in the number of triangles, and has an expected number of queries and runtime of
$O((\sqrt{n}\Delta(G)^{-1/6} + m^{3/4} \Delta(G)^{-1/2}) { \rm poly}(1/\epsilon)  )$, where $n$ is the number of nodes and $m$ is the number of edges. Under the same model, thanks to \cite[Lemma 48]{gilyen2019quantum}, we have an efficient algorithm for building a $(s, \poly\log (n/\epsilon), \epsilon)$-block encoding of the adjacency matrix $A$ of a graph, where $s$ is the sparsity of $A$. We use quantum access to the adjacency matrix of $G$ in the general graph model to build a block encoding of $A$. Then, we can estimate the number of triangles using the intuition behind eq.~\ref{eq:triangleadjacecny}.

The simple algorithm for computing the number of triangles is to build the block-encoding of $A^3$ and then estimate the trace. This leads to a runtime of $\widetilde{O}\left(\frac{\alpha^3 n \norm{A}}{\epsilon}\right)$ for an estimate with absolute error. However, we can obtain a much better runtime. The first trick is using the amplification of block-encoding and the second is using the algorithm for estimating the trace of $\Tr[M^TM]$. Observe that if we work with a normalized matrix $B=A/\|A\|$, if we obtain an estimate with a relative error of $Tr[B^3]$ we can obtain a relative estimate on $Tr[A^3] = \|A\|^3 \Tr[B^3]$.

\begin{cor}[Counting triangles - 1]
    \label{thm:counting triangles}
Let $\epsilon \in (0,1]$, and let $U_B$ be an $\alpha$-block-encoding of the normalized adjacency matrix $B=A/\|A\|$ of an undirected graph $G$ with adjacency matrix $A$.
There exists a quantum algorithm that computes a relative $\epsilon$ approximation of $\Delta(G)$ with high probability using $U_B$ and $U_B^\dagger$ for: 
$O\left( \frac{\alpha n}{\epsilon \Delta(G)} \log \left(\frac{n}{Tr[A]\epsilon}\right) \right).$
\end{cor}
\begin{proof}
    We use theorem~\ref{thm:prod of k amplified block-encodings} (product of k amplified block-encodings) to create a $(2^{1.5}, \frac{\epsilon \Tr[B]}{2n}$)-block-encoding of $B^3$ using $U_B$ and its inverse for $O( \alpha \log(\frac{n}{\Tr[B]\epsilon}))$ times. 
    We then use lemma~\ref{lem:quantum trace estimator relative} to estimate the trace of $\frac{B^3}{2^{1.5}}$ with relative error $\epsilon$ and using $U_B$ and its inverse for  $O\left( \frac{\alpha n}{\epsilon \Delta(G)} \log \left(\frac{n}{Tr[B]\epsilon}\right) \right).$ times.
\end{proof}

\begin{cor}[Counting triangles - 2 (normalized quantum access)]
    \label{thm:counting triangles - bound on spectral norm}
Let $\epsilon >0 $, and let $U_B$ be an $\alpha$-block-encoding of a normalized adjacency matrix $B = A/\|A\|$ of an undirected graph $G$.
There exists a quantum algorithm that computes a relative $\epsilon$-approximation of $\Delta(G)$ with high probability using $U_B$ and $U_B^\dagger$ for
$O\left(\frac{\sqrt{n}\alpha\kappa}{\epsilon \sqrt{\Delta[G]}} \log^3\left(\frac{\kappa\sqrt{n}}{\epsilon \sqrt{\Delta[G]}}\right) \right)$ times.
\end{cor}

\begin{proof}
Use lemma~\ref{lemma: BE of matrix power hermitian} to implement a $(2, O(\log\log(1/\epsilon)), \frac{\epsilon\sqrt{\Tr[B^3]}}{16\sqrt{2}\sqrt{n}})$-block-encoding of $B^{0.5}$ in time $O\left( \alpha \kappa \log^2\left(\frac{\kappa\sqrt{n}}{\epsilon \sqrt{\Tr[B^3]}}\right) \right)$ usages of $U_B$ and its inverse. Note that this won't affect the spectral norm of the block-encoding, as $\|B^{0.5}\|\leq 1$ if  $\|B\| \leq 1$. Now we combine the block-encodings of $B$ and $B^{0.5}$ with lemma~\ref{thm:product preamplified be 2022} (product of amplified block-encodings) to get a $(2\norm{B}\norm{B^{0.5}}, \frac{\epsilon\sqrt{Tr[B^3]}}{4\sqrt{n}})$-block-encoding of $B^{1.5}$ using $U_B$ and its inverse for 
$$O\left( \left( \alpha + \alpha\kappa \log^2\left(\frac{\kappa\sqrt{n}}{\epsilon \sqrt{\Tr[B^3]}}\right)  \right)\log \left( \frac{\sqrt{n}}{\epsilon\sqrt{\Tr[B^3]}} \right) \right).$$

Then, use trace estimation of lemma~\ref{corollary: trace estimation of product final} (trace estimation of product) to estimate of $Tr[B^{1.5 \dagger}B^{1.5}] = Tr[B^3]$
using $U_B$ and its inverse for 
$$O\left(\frac{\sqrt{n}}{\epsilon \sqrt{Tr[B^3]}}\left(\alpha\kappa \log^2\left(\frac{\kappa\sqrt{n}}{\epsilon \sqrt{\Tr[B^3]}}\right)  \right) \log \left( \frac{\sqrt{n}}{\epsilon\sqrt{\Tr[B^3]}} \right) \right).$$

\end{proof}

\subsection{Estimating the number of spanning trees}\label{sec:spanning trees}
It is well known that the problem of counting (or estimating) the number of spanning trees in a graph is related to the determinant of the Laplacian matrix of the graph (see Section~\ref{sec:vonneumannentropy} for an introduction to the formalism and definitions). There are many applications for efficient inference in tree mixture models~\cite{meila2000learning}, genomics, and network theory~\cite{wang2014clustering,kirby2016kirchhoff}.

Let $G$ be a connected, undirected graph of $n$
vertices. A spanning tree of $G$ is a sub-graph of $G$, which 
is a tree and includes all the vertices of $G$. We denote the number of spanning trees in $G$ as $t(G)$. The following theorem states that any principal minor of the Laplacian of $G$ where we remove any row and column in position $i \in [n]$, is equal to the number of spanning trees $t(G)$.

\begin{theorem}[{Kirchhoff Matrix-Tree Theorem \cite[Theorem 6.7]{kocay2016graphs} \cite{kirchhoff1847ueber}}]
Let $G$ be a connected graph without loops on $n$ vertices with Laplacian matrix $\mathcal{L}$. Let $\mathcal{L}(i|j)$ be the matrix by deleting the $i$-th row and $j$-th column of $\mathcal{L}$. Then for any $i \in [n]$, $t(G) = (-1)^{u+v} \det(\mathcal{L}(i|j))$
\end{theorem}

Let $\Delta_{\rm avg}, \Delta_{\rm max}$ 
be the average and maximum degrees of
vertices of the graph $G$, then the singular values of $L(i)$
lies in the interval $[1,2\Delta_{\rm max}-1]$~\cite{han} (see~\cite{zhang2011laplacian} for better bounds), and thus the condition number of $L$ is $\Theta(\Delta_{\rm max})$. It is shown in \cite[Corollary 5]{han} that a fast classical
algorithm to approximate $\log t(G)$ with relative error has complexity
$\widetilde{O}(n\Delta_{\rm max}/\epsilon^2\Delta_{\rm avg})$. While quantum algorithms for finding minimum spanning trees exist~\cite{heiligman2003quantum}, we are not aware of any quantum algorithm for \emph{counting} them.

We expect the quantum algorithm to return a good approximation of the number of spanning trees directly on the block-encoding of $L=\mathcal{L}/\|\mathcal{L}\|$, without removing any rows or columns. Define the matrix $\mathcal{L}(A|B)$ to be the matrix obtained removing rows indexed by $A \subset [n]$ and columns indexed by $B \subset [n]$ from $\mathcal{L}$. If our matrix is not symmetric, e.g. we have a block-encoding of $\mathcal{L}(i|j)$ for $j\neq i$, we might have to change some of our subroutines. For example, we can use an even function for the polynomial approximation of the logarithm (see appendix of~\cite{distributional}), instead of lemma~\ref{lemma:poly approx ln distributional} and using \cite[Corollary 18]{gilyen2019quantum} instead of theorem~\ref{thm:arbParity}.

\begin{cor}[Quantum algorithms for counting spanning trees]
\label{cor:spanning trees counting}
Let $G$ be a connected undirected graph, and suppose $U_L$ is an $\alpha$-block-encoding of $L(i|j)=\mathcal{L}(i|j)/\|\mathcal{L}(i|j)\|$. For any $\epsilon >0 $ there is a quantum algorithm that returns a relative error approximation of $t(G)$ with high probability in time $\widetilde{O}(n \alpha \Delta_{\rm max} /\epsilon)$.
\end{cor}
\begin{proof}
For ease of notation, we just denote as $\mathcal{L},L$ the matrices $\mathcal{L}(i|j), L(i|j)$. Observe that the proof of theorem~\ref{thm:logdet-svt} returns an absolute error $\epsilon/2$ for $\log\det(L)$ with a runtime of $\widetilde{O}_{\delta}(\frac{n\kappa\alpha}{\epsilon})$. Further observe that  

\begin{equation}\log\det(L) = \log\left( \prod_i \lambda_i / \|\mathcal{L}\|\right) = \log [\det(\mathcal{L})] - \log(\|\mathcal{L}\|^{n-1}).
\end{equation}
Hence, 
\begin{equation}|\log\det(\mathcal{L}) - \overline{\log\det(\mathcal{L})} | = |\log\det(L) - \overline{\log\det(L)} | \leq  \epsilon/2.
\end{equation}

Recall claim~\ref{claim:exponential-error} from the appendix. As $t(G) = e^{\log\det(\mathcal{L})}$, it follows that we can obtain                       
$\widetilde{t(G)}$ such that
$|\widetilde{t(G)} - t(G)| \leq t(G)\epsilon$.
\end{proof}

\subsection{Estimating the effective resistance}\label{sec:effective resistance}

Besides having applications in machine learning, one reason 
for to the importance of estimating the number of 
spanning trees in a graph is its relation to the effective
resistance. We consider a weighted graph $G=(V,E)$ on $n$ nodes with edge weights $w_{ij} \in \mathbb{R}^+$ for edge $(i,j) \in E$. Recall that we denote by $\mathcal{L}(i,j)$ the $(n-2) \times (n-2)$ matrix obtained from the Laplacian of $G$ by removing the rows and columns associated with vertices $i$ and $j$. 
The following statement, which we can take as a definition of effective resistance, is actually a theorem (see~\cite[Theorem 2.10]{vos2016methods} and~\cite{bapat2003simple}). 

\begin{definition}[Effective resistance\cite{vos2016methods}] Let $G=(V,E)$ be a simple weighted connected graph with $|V|\geq 3$, and with weights in $\mathbb{R}^+$ for edge $(i,j) \in E$, where $w_{ij}=0$ if $(i,j)\neq E$. The resistance of edge $e=(i,j)$ is $R_e=\frac{1}{w_{ij}}$. The effective resistance between any pair of vertices $i,j \in V$ is
\be
R_{i,j} := \frac{\det \mathcal{L}(i,j)}{\det \mathcal{L}(i)}.
\ee
\end{definition}

Quantum algorithms to compute
$R_{i,j}$ have been studied in~\cite{GuomingWang,ItoJeffery}. For instance, within the Belovs model of quantum random walk,  it is possible to estimate the effective resistance between two nodes with a total runtime of $O(\sqrt{R_{i,j}W}/\epsilon^2)$, where $W=\sum_{i,j \in E} w_{ij}$ is the total weight of the graph ~\cite{apers2019unified,piddock2019quantum}.  In the block-encoding model, we can use the following result.

\begin{thm}[Quantum algorithm for estimating the effective resistance]
Let $G$ be a simple weighted connected graph, and suppose $U_{L_1}$ is an $\alpha_1$-block-encoding of $L(i)=\mathcal{L}(i)/\|\mathcal{L}(i)\|$ and $U_{L_2}$ is an $\alpha_2$-block-encoding of $L(i,j)=\mathcal{L}(i,j)/\|\mathcal{L}(i,j)\|$. For any $\epsilon > 0$ and $i,j \in V$, there is a quantum algorithm that returns a relative error approximation of $R_{i,j}$ with high probability in time $\widetilde{O}(n \alpha \kappa(\mathcal{L})/\epsilon)$, where $\alpha = \max \{\alpha_1, \alpha_2\}$.
\end{thm}
\begin{proof}
First, observe that $\kappa(\mathcal{L}(i,j)) \leq \kappa(\mathcal{L}(i)) \leq \kappa(\mathcal{L})$ (see lemma~\ref{lem:martingaleobservation} in appendix~\ref{appx:polyapprox}). The effective resistance between $a$ and $b$ can be obtained by running twice the quantum algorithm for the log-determinant for the numerator and the denominator, with absolute error $\epsilon/4$. In fact, 
\begin{align}
R_{i,j} & = \frac{e^{\log\det \mathcal{L}(i,j)}}{e^{\log\det \mathcal{L}(i)}} = e^{\log\det \mathcal{L}(i,j) -  \log\det \mathcal{L}(i) } \nonumber \\
 & = e^{\log\det L(i,j) -  \log\det L(i) -\log\left(\|\mathcal{L}(i)\|^{(n-1)}\right) + \log\left(\|\mathcal{L}(i,j)\|^{(n-2)}\right)}
\end{align}

Consider running the algorithm of theorem~\ref{thm:logdet-svt} to obtain an absolute error $\epsilon/4$ (this will cost $\widetilde{O}(\frac{n\alpha\kappa}{\epsilon})$ queries to $U_{L_1}$ and $U_{L_2}$, so to get
$ \overline{\log\det  L(i,j)}$ and 
$\overline{\log\det  L(i)}$. Then similarly to what we observed before,
$$\left|\Big(\log\det  \mathcal{L}(i,j) - \log\det  \mathcal{L}(i)\Big) - \Big(\overline{\log\det  \mathcal{L}(i,j)}- \overline{\log\det  \mathcal{L}(i)}\Big)\right| \leq \frac{\epsilon}{2}$$

Concluding, using the knowledge of the spectral norms of the operators, we can use claim~\ref{claim:exponential-error} to get
$|R_{i,j} - \overline{R_{i,j}}| \leq R_{i,j} \epsilon $

\end{proof}

\section{Conclusions and future works}
There are few striking differences between quantum and classical algorithms. Some polynomial approximations that are used in classical algorithms (like Gaussian quadrature, the Lanczos method, and the Cauchy integral~\cite{dong2017scalable, dorn2015stochastic, hale2008computing}) leads to small polynomial dependence in the condition number. Future research will explore how to leverage these polynomial approximations within the quantum setting.  There is a flurry of classical techniques used for stochastic trace estimation~\cite{Avron}, which seems not to lead to any advantage for the quantum case. In fact, all of our quantum algorithms have a runtime dependence that is linear in the approximation error, while most of the classical algorithms have a quadratic dependence. Interestingly, a few recent classical algorithms that use new stochastic trace estimators~\cite{meyer2021hutch++} have a linear dependence in the error. This leaves the question open whether quantum algorithms can beat classical algorithms and be sub-linear in the approximation error. The authors are not aware of any lower bounds for the trace estimation problem.

We believe the quantum algorithm for the trace of the inverse can be improved, as it is polynomially slower than the classical algorithm in some parameters.  A possible generalization of the quantum algorithm for von Neumann entropy is a quantum algorithm for estimating the relative entropy between two quantum states. Interestingly, this algorithm can be linked again to the number of spanning trees of a graph, as it was recently shown in~\cite{giovannetti2013kirchhoff}, paving the way for different algorithms for this problem, with different runtimes than what we discussed in Section~\ref{sec:applications}. Many classical algorithms also report the variance associated with these estimators. While the variance of the estimator proposed in this work is derived directly from the variance of the subroutines we employ, the exploration of more advanced amplitude estimation algorithms~\cite{cornelissen2023sublinear}, promising improved variance dependency, is deferred to future research. We leave for future work the study of algorithms for estimating the distance between PSD matrices (which can be reduced to a spectral sum~\cite{bhatia2019bures}) and computing the log-determinant of non-square matrices. Future works include also the study of more applications of quantum algorithms for spectral sums, and more numerical experiments (e.g. truncating the condition number for the triangle counting algorithm) and other algorithms for counting the number of triangles (for example using~\cite{gall2022quantum}).

\section*{Acknowledgments}
We would like to thank, in sparse order,
Chris  Cade,
Andr{\'a}s Gily{\'e}n, 
Sander Gribling, 
Yassine Hamoudi, 
Iordanis Kerenidis,
Pablo Rotondo,
Ashley Montanaro, 
Armando Bellante,
Varun Narasimhachar,
Stephen Piddock,
Sathyawageeswar Subramanian, 
Liron Mor Yosef, 
Miklos Santha, and
Haim Avron
for useful discussions and feedback on the manuscript. AL and CS have been supported by QuantERA ERA-NET Cofund in Quantum Technologies implemented within the European
Union's Horizon 2020 Programme (QuantAlgo project). 
AL has been also supported by ANTR and CS is supported by EPSRC grants EP/L021005/1 and EP/R043957/1. This work was supported by the National Research Foundation, Singapore, and A*STAR under its CQT Bridging Grant. We also acknowledge funding from the Quantum Engineering Programme (QEP 2.0) under grants NRF2021-QEP2-02-P05 and NRF2021-QEP2-02-P01.

\section*{Contribution}
AL and CS conceived the initial ideas and an early version of the manuscript. AL wrote the final version of the manuscript.

\bibliography{sn-bibliography}%

\begin{thebibliography}{100}

\bibitem{aaronson2020quantum}
Scott Aaronson and Patrick Rall.
\newblock Quantum approximate counting, simplified.
\newblock In {\em Symposium on simplicity in algorithms}, pages 24--32. SIAM, 2020.

\bibitem{ahmad2021skeleton}
Tasweer Ahmad, Lianwen Jin, Luojun Lin, and GuoZhi Tang.
\newblock Skeleton-based action recognition using sparse spatio-temporal gcn with edge effective resistance.
\newblock {\em Neurocomputing}, 423:389--398, 2021.

\bibitem{aho1974design}
Alfred~V Aho and John~E Hopcroft.
\newblock {\em The design and analysis of computer algorithms}.
\newblock Pearson Education India, 1974.

\bibitem{alev2017graph}
Vedat~Levi Alev, Nima Anari, Lap~Chi Lau, and Shayan~Oveis Gharan.
\newblock Graph clustering using effective resistance.
\newblock {\em arXiv preprint arXiv:1711.06530}, 2017.

\bibitem{allcock2023constant}
Jonathan Allcock, Jinge Bao, Jo{\~a}o~F Doriguello, Alessandro Luongo, and Miklos Santha.
\newblock Constant-depth circuits for uniformly controlled gates and boolean functions with application to quantum memory circuits.
\newblock {\em arXiv preprint arXiv:2308.08539}, 2023.

\bibitem{alter2000singular}
Orly Alter, Patrick~O Brown, and David Botstein.
\newblock Singular value decomposition for genome-wide expression data processing and modeling.
\newblock {\em Proceedings of the National Academy of Sciences}, 97(18):10101--10106, 2000.

\bibitem{ambainis2000computing}
Andris Ambainis, Leonard~J Schulman, and Umesh~V Vazirani.
\newblock Computing with highly mixed states.
\newblock In {\em Proceedings of the thirty-second annual ACM symposium on Theory of computing}, pages 697--704, 2000.

\bibitem{apers2019unified}
Simon Apers, Andr{\'a}s Gily{\'e}n, and Stacey Jeffery.
\newblock A unified framework of quantum walk search.
\newblock {\em arXiv preprint arXiv:1912.04233}, 2019.

\bibitem{aune}
Erlend Aune, Daniel~P Simpson, and Jo~Eidsvik.
\newblock Parameter estimation in high dimensional gaussian distributions.
\newblock {\em Statistics and Computing}, 24(2):247--263, 2014.

\bibitem{avron2011randomized}
Haim Avron and Sivan Toledo.
\newblock Randomized algorithms for estimating the trace of an implicit symmetric positive semi-definite matrix.
\newblock {\em Journal of the ACM (JACM)}, 58(2):1--34, 2011.

\bibitem{Avron}
Haim Avron and Sivan Toledo.
\newblock Randomized algorithms for estimating the trace of an implicit symmetric positive semi-definite matrix.
\newblock {\em Journal of the ACM (JACM)}, 58(2):1--34, 2011.

\bibitem{ayres2000sisporto}
Diogo Ayres-de Campos, Joao Bernardes, Antonio Garrido, Joaquim Marques-de Sa, and Luis Pereira-Leite.
\newblock Sisporto 2.0: a program for automated analysis of cardiotocograms.
\newblock {\em Journal of Maternal-Fetal Medicine}, 9(5):311--318, 2000.

\bibitem{bai1996bounds}
Zhaojun Bai and Gene~H Golub.
\newblock Bounds for the trace of the inverse and the determinant of symmetric positive definite matrices.
\newblock {\em Annals of Numerical Mathematics}, 4:29--38, 1996.

\bibitem{banerjee2014feature}
Monami Banerjee and Nikhil~R Pal.
\newblock Feature selection with svd entropy: Some modification and extension.
\newblock {\em Information Sciences}, 264:118--134, 2014.

\bibitem{bapat1996laplacian}
Ravindra~B Bapat.
\newblock The laplacian matrix of a graph.
\newblock {\em Mathematics Student-India}, 65(1):214--223, 1996.

\bibitem{bapat2003simple}
Ravindra~B Bapat, Ivan Gutmana, and Wenjun Xiao.
\newblock A simple method for computing resistance distance.
\newblock {\em Zeitschrift f{\"u}r Naturforschung A}, 58(9-10):494--498, 2003.

\bibitem{barry1999monte}
Ronald~Paul Barry and R~Kelley Pace.
\newblock Monte carlo estimates of the log determinant of large sparse matrices.
\newblock {\em Linear Algebra and its applications}, 289(1-3):41--54, 1999.

\bibitem{bengtsson2017geometry}
Ingemar Bengtsson and Karol {\.Z}yczkowski.
\newblock {\em Geometry of quantum states: an introduction to quantum entanglement}.
\newblock Cambridge university press, Cambridge, 2017.

\bibitem{bhatia2019bures}
Rajendra Bhatia, Tanvi Jain, and Yongdo Lim.
\newblock On the bures--wasserstein distance between positive definite matrices.
\newblock {\em Expositiones Mathematicae}, 37(2):165--191, 2019.

\bibitem{Boutsidis}
Christos Boutsidis, Petros Drineas, Prabhanjan Kambadur, Eugenia-Maria Kontopoulou, and Anastasios Zouzias.
\newblock A randomized algorithm for approximating the log determinant of a symmetric positive definite matrix.
\newblock {\em Linear Algebra and its Applications}, 533:95--117, 2017.

\bibitem{brassard2002quantum}
Gilles Brassard, Peter Hoyer, Michele Mosca, and Alain Tapp.
\newblock Quantum amplitude amplification and estimation.
\newblock {\em Contemporary Mathematics}, 305:53--74, 2002.

\bibitem{braunstein2006laplacian}
Samuel~L Braunstein, Sibasish Ghosh, and Simone Severini.
\newblock The laplacian of a graph as a density matrix: a basic combinatorial approach to separability of mixed states.
\newblock {\em Annals of Combinatorics}, 10:291--317, 2006.

\bibitem{Cade}
Chris Cade and Ashley Montanaro.
\newblock The quantum complexity of computing schatten p-norms.
\newblock In {\em 13th Conference on the Theory of Quantum Computation, Communication and Cryptography (TQC 2018)}. Schloss Dagstuhl-Leibniz-Zentrum fuer Informatik, 2018.

\bibitem{camps2022explicit}
Daan Camps, Lin Lin, Roel Van~Beeumen, and Chao Yang.
\newblock Explicit quantum circuits for block encodings of certain sparse matrice.
\newblock {\em arXiv preprint arXiv:2203.10236}, 2022.

\bibitem{caraiani2014predictive}
Petre Caraiani.
\newblock The predictive power of singular value decomposition entropy for stock market dynamics.
\newblock {\em Physica A: Statistical Mechanics and its Applications}, 393:571--578, 2014.

\bibitem{chakraborty2018power}
S.~Chakraborty, A.~Gily{\'e}n, and S.~Jeffery.
\newblock The power of block-encoded matrix powers: improved regression techniques via faster hamiltonian simulation.
\newblock In {\em Proc. 46\textsuperscript{th} International Colloquium on Automata, Languages, and Programming}, pages 33:1--33:14, 2019.

\bibitem{chakraborty2022quantum}
Shantanav Chakraborty, Aditya Morolia, and Anurudh Peduri.
\newblock Quantum regularized least squares.
\newblock {\em arXiv preprint arXiv:2206.13143}, 2022.

\bibitem{chao2020finding}
Rui Chao, Dawei Ding, Andras Gilyen, Cupjin Huang, and Mario Szegedy.
\newblock Finding angles for quantum signal processing with machine precision.
\newblock {\em arXiv preprint arXiv:2003.02831}, 2020.

\bibitem{chen2023krylov}
Tyler Chen and Eric Hallman.
\newblock Krylov-aware stochastic trace estimation.
\newblock {\em SIAM Journal on Matrix Analysis and Applications}, 44(3):1218--1244, 2023.

\bibitem{chen2022randomized}
Tyler Chen, Thomas Trogdon, and Shashanka Ubaru.
\newblock Randomized matrix-free quadrature for spectrum and spectral sum approximation.
\newblock {\em arXiv preprint arXiv:2204.01941}, 2022.

\bibitem{childs2017quantum}
Andrew~M Childs, Robin Kothari, and Rolando~D Somma.
\newblock Quantum algorithm for systems of linear equations with exponentially improved dependence on precision.
\newblock {\em SIAM Journal on Computing}, 46(6):1920--1950, 2017.

\bibitem{chowdhury2021computing}
Anirban~N Chowdhury, Rolando~D Somma, and Yi{\u{g}}it Suba{\c{s}}{\i}.
\newblock Computing partition functions in the one-clean-qubit model.
\newblock {\em Physical Review A}, 103(3):032422, 2021.

\bibitem{cichoki2014bregman}
Andrzej Cichocki, Sergio Cruces, and Shun-ichi Amari.
\newblock Log-determinant divergences revisited: Alpha-beta and gamma log-det divergences.
\newblock {\em Entropy}, 17(5):2988--3034, 2015.

\bibitem{cornelissen2023sublinear}
Arjan Cornelissen and Yassine Hamoudi.
\newblock A sublinear-time quantum algorithm for approximating partition functions.
\newblock In {\em Proceedings of the 2023 Annual ACM-SIAM Symposium on Discrete Algorithms (SODA)}, pages 1245--1264. SIAM, 2023.

\bibitem{dashti2011uncertainty}
Masoumeh Dashti and Andrew~M Stuart.
\newblock Uncertainty quantification and weak approximation of an elliptic inverse problem.
\newblock {\em SIAM Journal on Numerical Analysis}, 49(6):2524--2542, 2011.

\bibitem{datta2008quantum}
Animesh Datta, Anil Shaji, and Carlton~M Caves.
\newblock Quantum discord and the power of one qubit.
\newblock {\em Physical Review Letters}, 100(5):050502, 2008.

\bibitem{datta2007role}
Animesh Datta and Guifre Vidal.
\newblock Role of entanglement and correlations in mixed-state quantum computation.
\newblock {\em Physical Review A}, 75(4):042310, 2007.

\bibitem{dong2017scalable}
Kun Dong, David Eriksson, Hannes Nickisch, David Bindel, and Andrew~G Wilson.
\newblock Scalable log determinants for gaussian process kernel learning.
\newblock In {\em Advances in Neural Information Processing Systems}, pages 6327--6337, 2017.

\bibitem{dong2021efficient}
Yulong Dong, Xiang Meng, K~Birgitta Whaley, and Lin Lin.
\newblock Efficient phase-factor evaluation in quantum signal processing.
\newblock {\em Physical Review A}, 103(4):042419, 2021.

\bibitem{doriguello2022quantum}
Jo{\~a}o~F Doriguello, Alessandro Luongo, Jinge Bao, Patrick Rebentrost, and Miklos Santha.
\newblock Quantum algorithm for stochastic optimal stopping problems with applications in finance.
\newblock {\em Leibniz Int. Proc. Inf.}, 232(arXiv: 2111.15332):2--1, 2022.

\bibitem{dorn2015stochastic}
Sebastian Dorn and Torsten~A En{\ss}lin.
\newblock Stochastic determination of matrix determinants.
\newblock {\em Physical Review E}, 92(1):013302, 2015.

\bibitem{easley2012networks}
David Easley, Jon Kleinberg, et~al.
\newblock Networks, crowds, and markets: Reasoning about a highly connected world.
\newblock {\em Significance}, 9:43--44, 2012.

\bibitem{epperly2024xtrace}
Ethan~N Epperly, Joel~A Tropp, and Robert~J Webber.
\newblock Xtrace: Making the most of every sample in stochastic trace estimation.
\newblock {\em SIAM Journal on Matrix Analysis and Applications}, 45(1):1--23, 2024.

\bibitem{fortunato2010community}
Santo Fortunato.
\newblock Community detection in graphs.
\newblock {\em Physics reports}, 486(3-5):75--174, 2010.

\bibitem{fujii2018impossibility}
Keisuke Fujii, Hirotada Kobayashi, Tomoyuki Morimae, Harumichi Nishimura, Shuhei Tamate, and Seiichiro Tani.
\newblock Impossibility of classically simulating one-clean-qubit model with multiplicative error.
\newblock {\em Physical Review Letters}, 120(20):200502, 2018.

\bibitem{gall2022quantum}
Fran{\c{c}}ois~Le Gall and Iu-Iong Ng.
\newblock Quantum approximate counting for markov chains and application to collision counting.
\newblock {\em arXiv preprint arXiv:2204.02552}, 2022.

\bibitem{distributional}
Andr{\'a}s Gily{\'e}n and Tongyang Li.
\newblock Distributional property testing in a quantum world.
\newblock In {\em 11th Innovations in Theoretical Computer Science Conference (ITCS 2020)}. Schloss Dagstuhl-Leibniz-Zentrum f{\"u}r Informatik, 2020.

\bibitem{gilyen2019quantum}
Andr{\'a}s Gily{\'e}n, Yuan Su, Guang~Hao Low, and Nathan Wiebe.
\newblock Quantum singular value transformation and beyond: exponential improvements for quantum matrix arithmetics.
\newblock In {\em Proceedings of the 51st Annual ACM SIGACT Symposium on Theory of Computing}, pages 193--204, 2019.

\bibitem{giovannetti2008architectures}
Vittorio Giovannetti, Seth Lloyd, and Lorenzo Maccone.
\newblock Architectures for a quantum random access memory.
\newblock {\em Physical Review A}, 78(5):052310, 2008.

\bibitem{giovannetti2013kirchhoff}
Vittorio Giovannetti and Simone Severini.
\newblock Kirchhoff's matrix-tree theorem revisited: Counting spanning trees with the quantum relative entropy.
\newblock {\em Advances in Network Complexity}, pages 177--190, 2013.

\bibitem{giurgica2022low}
Tudor Giurgica-Tiron, Iordanis Kerenidis, Farrokh Labib, Anupam Prakash, and William Zeng.
\newblock Low depth algorithms for quantum amplitude estimation.
\newblock {\em Quantum}, 6:745, 2022.

\bibitem{goktacs2020benchmarking}
Oktay G{\"o}kta{\c{s}}, Weng~Kian Tham, Kent Bonsma-Fisher, and Aharon Brodutch.
\newblock Benchmarking quantum processors with a single qubit.
\newblock {\em Quantum Information Processing}, 19(5):1--17, 2020.

\bibitem{golub1997generalized}
Gene~H Golub and Urs Von~Matt.
\newblock Generalized cross-validation for large-scale problems.
\newblock {\em Journal of Computational and Graphical Statistics}, 6(1):1--34, 1997.

\bibitem{thesissander}
Sander Gribling.
\newblock {\em Applications of optimization to factorization ranks and quantum information theory}.
\newblock PhD thesis, Tilburg University, 2019.

\bibitem{gribling2021improving}
Sander Gribling, Iordanis Kerenidis, and D{\'a}niel Szil{\'a}gyi.
\newblock Improving quantum linear system solvers via a gradient descent perspective.
\newblock {\em arXiv preprint arXiv:2109.04248}, 2021.

\bibitem{grinko2021iterative}
Dmitry Grinko, Julien Gacon, Christa Zoufal, and Stefan Woerner.
\newblock Iterative quantum amplitude estimation.
\newblock {\em npj Quantum Information}, 7(1):52, 2021.

\bibitem{gur2021sublinear}
Tom Gur, Min-Hsiu Hsieh, and Sathyawageeswar Subramanian.
\newblock Sublinear quantum algorithms for estimating von neumann entropy.
\newblock {\em arXiv preprint arXiv:2111.11139}, 2021.

\bibitem{gutman2001energy}
Ivan Gutman.
\newblock The energy of a graph: old and new results, algebraic combinatorics and applications.
\newblock {\em Betten, A., Kohner, A., Laue, R., Wassermann, A.(eds.)}, pages 196--211, 2001.

\bibitem{hale2008computing}
Nicholas Hale, Nicholas~J Higham, and Lloyd~N Trefethen.
\newblock {Computing $A^\alpha, \log(A)$, and related matrix functions by contour integrals}.
\newblock {\em SIAM Journal on Numerical Analysis}, 46(5):2505--2523, 2008.

\bibitem{hamoudi2018quantum}
Yassine Hamoudi and Fr{\'e}d{\'e}ric Magniez.
\newblock Quantum chebyshev's inequality and applications.
\newblock {\em arXiv preprint arXiv:1807.06456}, 2018.

\bibitem{han2017approximating}
Insu Han, Dmitry Malioutov, Haim Avron, and Jinwoo Shin.
\newblock Approximating spectral sums of large-scale matrices using stochastic chebyshev approximations.
\newblock {\em SIAM Journal on Scientific Computing}, 39(4):A1558--A1585, 2017.

\bibitem{han}
Insu Han, Dmitry Malioutov, and Jinwoo Shin.
\newblock Large-scale log-determinant computation through stochastic chebyshev expansions.
\newblock In {\em International Conference on Machine Learning}, pages 908--917, 2015.

\bibitem{han2012graph}
Lin Han, Francisco Escolano, Edwin~R Hancock, and Richard~C Wilson.
\newblock Graph characterizations from von neumann entropy.
\newblock {\em Pattern Recognition Letters}, 33(15):1958--1967, 2012.

\bibitem{han2022adbench}
Songqiao Han, Xiyang Hu, Hailiang Huang, Minqi Jiang, and Yue Zhao.
\newblock Adbench: Anomaly detection benchmark.
\newblock {\em Advances in Neural Information Processing Systems}, 35:32142--32159, 2022.

\bibitem{hann2021resilience}
Connor~T Hann, Gideon Lee, SM~Girvin, and Liang Jiang.
\newblock Resilience of quantum random access memory to generic noise.
\newblock {\em PRX Quantum}, 2(2):020311, 2021.

\bibitem{hardt2012simple}
Moritz Hardt, Katrina Ligett, and Frank McSherry.
\newblock A simple and practical algorithm for differentially private data release.
\newblock In {\em Advances in Neural Information Processing Systems}, pages 2339--2347, 2012.

\bibitem{heiligman2003quantum}
Mark Heiligman.
\newblock Quantum algorithms for lowest weight paths and spanning trees in complete graphs.
\newblock {\em arXiv preprint quant-ph/0303131}, 2003.

\bibitem{horn2012matrix}
Roger~A Horn and Charles~R Johnson.
\newblock {\em Matrix analysis}.
\newblock Cambridge university press, Cambridge, 2012.

\bibitem{ItoJeffery}
Tsuyoshi Ito and Stacey Jeffery.
\newblock Approximate span programs.
\newblock {\em Algorithmica}, 81(6):2158--2195, 2019.

\bibitem{jaques2023qram}
Samuel Jaques and Arthur~G Rattew.
\newblock Qram: A survey and critique.
\newblock {\em arXiv preprint arXiv:2305.10310}, 2023.

\bibitem{powering}
Mark~R Jerrum, Leslie~G Valiant, and Vijay~V Vazirani.
\newblock Random generation of combinatorial structures from a uniform distribution.
\newblock {\em Theoretical computer science}, 43:169--188, 1986.

\bibitem{jiang2021optimal}
Shuli Jiang, Hai Pham, David Woodruff, and Richard Zhang.
\newblock Optimal sketching for trace estimation.
\newblock {\em Advances in Neural Information Processing Systems}, 34:23741--23753, 2021.

\bibitem{jones2024controlling}
Jonathan~A Jones.
\newblock Controlling nmr spin systems for quantum computation.
\newblock {\em Progress in Nuclear Magnetic Resonance Spectroscopy}, 2024.

\bibitem{jordan2008quantum}
Stephen~Paul Jordan.
\newblock {\em Quantum computation beyond the circuit model}.
\newblock PhD thesis, Massachusetts Institute of Technology, 2008.

\bibitem{karimi2023power}
Mahsa Karimi, Ali Javadi-Abhari, Christoph Simon, and Roohollah Ghobadi.
\newblock The power of one clean qubit in supervised machine learning.
\newblock {\em Scientific Reports}, 13(1):19975, 2023.

\bibitem{katz1953new}
Leo Katz.
\newblock A new status index derived from sociometric analysis.
\newblock {\em Psychometrika}, 18(1):39--43, 1953.

\bibitem{kerenidis2020quantumEM}
Iordanis Kerenidis, Alessandro Luongo, and Anupam Prakash.
\newblock Quantum expectation-maximization for gaussian mixture models.
\newblock In {\em International Conference on Machine Learning}, pages 5187--5197. PMLR, 2020.

\bibitem{kerenidis2017recommendation}
Iordanis Kerenidis and Anupam Prakash.
\newblock Quantum recommendation systems.
\newblock In {\em 8th Innovations in Theoretical Computer Science Conference (ITCS 2017)}. Schloss Dagstuhl-Leibniz-Zentrum fuer Informatik, 2017.

\bibitem{kerenidis2020quantum}
Iordanis Kerenidis and Anupam Prakash.
\newblock Quantum gradient descent for linear systems and least squares.
\newblock {\em Physical Review A}, 101(2):022316, 2020.

\bibitem{kirby2016kirchhoff}
Edward~C Kirby, Roger~B Mallion, Paul Pollak, and Pawe{\l}~J Skrzy{\'n}ski.
\newblock What kirchhoff actually did concerning spanning trees in electrical networks and its relationship to modern graph-theoretical work.
\newblock {\em Croatica Chemica Acta}, 89(4):403--417, 2016.

\bibitem{kirchhoff1847ueber}
Gustav Kirchhoff.
\newblock Ueber die aufl{\"o}sung der gleichungen, auf welche man bei der untersuchung der linearen vertheilung galvanischer str{\"o}me gef{\"u}hrt wird.
\newblock {\em Annalen der Physik}, 148(12):497--508, 1847.

\bibitem{kittaneh1985inequalities}
Fuad Kittaneh.
\newblock Inequalities for the schatten p-norm.
\newblock {\em Glasgow Mathematical Journal}, 26(2):141--143, 1985.

\bibitem{knill1998power}
Emanuel Knill and Raymond Laflamme.
\newblock Power of one bit of quantum information.
\newblock {\em Physical Review Letters}, 81(25):5672, 1998.

\bibitem{kocay2016graphs}
William Kocay and Donald~L Kreher.
\newblock {\em Graphs, algorithms, and optimization}.
\newblock CRC Press, 2016.

\bibitem{gramarandomized}
Eugenia-Maria Kontopoulou, Gregory-Paul Dexter, Wojciech Szpankowski, Ananth Grama, and Petros Drineas.
\newblock Randomized linear algebra approaches to estimate the von neumann entropy of density matrices.
\newblock {\em arXiv preprint arXiv:1801.01072v3}, 2018.

\bibitem{lecun1998mnist}
Yann LeCun.
\newblock The mnist database of handwritten digits.
\newblock {\em http://yann. lecun. com/exdb/mnist/}, 1998.

\bibitem{li2010making}
Mu~Li, James Tin-Yau Kwok, and Baoliang L{\"u}.
\newblock Making large-scale nystr{\"o}m approximation possible.
\newblock In {\em Proceedings of the 27th International Conference on Machine Learning, ICML 2010}, page 631, 2010.

\bibitem{li2018quantum}
Tongyang Li and Xiaodi Wu.
\newblock Quantum query complexity of entropy estimation.
\newblock {\em IEEE Transactions on Information Theory}, 65(5):2899--2921, 2018.

\bibitem{lohweg2012banknote}
Volker Lohweg and H~Doerksen.
\newblock Banknote authentication data set, 2012.

\bibitem{low2019hamiltonian}
Guang~Hao Low and Isaac~L Chuang.
\newblock Hamiltonian simulation by qubitization.
\newblock {\em Quantum}, 3:163, 2019.

\bibitem{lu2015individualized}
Yu~Lu and Sahand~N Negahban.
\newblock Individualized rank aggregation using nuclear norm regularization.
\newblock In {\em 2015 53rd Annual Allerton Conference on Communication, Control, and Computing (Allerton)}, pages 1473--1479. IEEE, 2015.

\bibitem{majumdar2011algorithm}
Angshul Majumdar and Rabab~K Ward.
\newblock An algorithm for sparse mri reconstruction by schatten $p$-norm minimization.
\newblock {\em Magnetic resonance imaging}, 29(3):408--417, 2011.

\bibitem{maronese2023quantum}
Marco Maronese, Massimiliano Incudini, Luca Asproni, and Enrico Prati.
\newblock The quantum amplitude estimation algorithms on near-term devices: A practical guide.
\newblock {\em Quantum Reports}, 6(1):1--13, 2023.

\bibitem{meila2000learning}
Marina Meila and Michael~I Jordan.
\newblock Learning with mixtures of trees.
\newblock {\em Journal of Machine Learning Research}, 1(Oct):1--48, 2000.

\bibitem{meyer2021hutch++}
Raphael~A Meyer, Cameron Musco, Christopher Musco, and David~P Woodruff.
\newblock Hutch++: Optimal stochastic trace estimation.
\newblock In {\em Symposium on Simplicity in Algorithms (SOSA)}, pages 142--155. SIAM, 2021.

\bibitem{Minello}
Giorgia Minello, Luca Rossi, and Andrea Torsello.
\newblock {On the von Neumann entropy of graphs}.
\newblock {\em Journal of Complex Networks}, 7(4):491--514, 2018.

\bibitem{montanaro2015quantum}
Ashley Montanaro.
\newblock {Quantum speedup of Monte Carlo methods}.
\newblock {\em Proceedings of the Royal Society A: Mathematical, Physical and Engineering Sciences}, 471(2181):20150301, 2015.

\bibitem{morimae2014hardness}
Tomoyuki Morimae, Keisuke Fujii, and Joseph~F Fitzsimons.
\newblock Hardness of classically simulating the one-clean-qubit model.
\newblock {\em Physical Review Letters}, 112(13):130502, 2014.

\bibitem{motlagh2023generalized}
Danial Motlagh and Nathan Wiebe.
\newblock Generalized quantum signal processing.
\newblock {\em arXiv preprint arXiv:2308.01501}, 2023.

\bibitem{musco2017spectrum}
Cameron Musco, Praneeth Netrapalli, Aaron Sidford, Shashanka Ubaru, and David~P Woodruff.
\newblock Spectrum approximation beyond fast matrix multiplication: Algorithms and hardness.
\newblock {\em arXiv preprint arXiv:1704.04163}, 2017.

\bibitem{nakatsukasa2020fast}
Yuji Nakatsukasa.
\newblock Fast and stable randomized low-rank matrix approximation.
\newblock {\em arXiv preprint arXiv:2009.11392}, 2020.

\bibitem{netrapalli2014non}
Praneeth Netrapalli, UN~Niranjan, Sujay Sanghavi, Animashree Anandkumar, and Prateek Jain.
\newblock {Non-convex robust PCA}.
\newblock In {\em Advances in Neural Information Processing Systems}, pages 1107--1115, 2014.

\bibitem{nie2012low}
Feiping Nie, Heng Huang, and Chris Ding.
\newblock Low-rank matrix recovery via efficient schatten p-norm minimization.
\newblock In {\em Twenty-sixth AAAI conference on artificial intelligence}, 2012.

\bibitem{nielsen2002quantum}
Michael~A Nielsen and Isaac Chuang.
\newblock {\em Quantum Computation and Quantum Information: 10th Anniversary Edition}.
\newblock Cambridge University Press, Cambridge, 2010.

\bibitem{pace2004}
R~Kelley Pace and James~P LeSage.
\newblock Chebyshev approximation of log-determinants of spatial weight matrices.
\newblock {\em Computational Statistics \& Data Analysis}, 45(2):179--196, 2004.

\bibitem{pang2019deep}
Guansong Pang, Chunhua Shen, and Anton Van Den~Hengel.
\newblock Deep anomaly detection with deviation networks.
\newblock In {\em Proceedings of the 25th ACM SIGKDD international conference on knowledge discovery \& data mining}, pages 353--362, 2019.

\bibitem{passante2012experimental}
Gina Passante.
\newblock {\em On experimental deterministic quantum computation with one quantum bit (DQC1)}.
\newblock Phd thesis, University of Waterloo, 2012.

\bibitem{passerini2008neumann}
Filippo Passerini and Simone Severini.
\newblock The von neumann entropy of networks.
\newblock {\em Available at SSRN 1382662}, 2008.

\bibitem{persson2022improved}
David Persson, Alice Cortinovis, and Daniel Kressner.
\newblock Improved variants of the hutch++ algorithm for trace estimation.
\newblock {\em SIAM Journal on Matrix Analysis and Applications}, 43(3):1162--1185, 2022.

\bibitem{persson2023randomized}
David Persson and Daniel Kressner.
\newblock Randomized low-rank approximation of monotone matrix functions.
\newblock {\em SIAM Journal on Matrix Analysis and Applications}, 44(2):894--918, 2023.

\bibitem{piddock2019quantum}
Stephen Piddock.
\newblock Quantum walk search algorithms and effective resistance.
\newblock {\em arXiv preprint arXiv:1912.04196}, 2019.

\bibitem{preskill1998lecture}
John Preskill.
\newblock Lecture notes for physics 229: Quantum information and computation.
\newblock {\em California Institute of Technology}, 16(1):1--8, 1998.

\bibitem{quek2024multivariate}
Yihui Quek, Eneet Kaur, and Mark~M Wilde.
\newblock Multivariate trace estimation in constant quantum depth.
\newblock {\em Quantum}, 8:1220, 2024.

\bibitem{rashtchian2020vector}
Cyrus Rashtchian, David~P Woodruff, and Hanlin Zhu.
\newblock Vector-matrix-vector queries for solving linear algebra, statistics, and graph problems.
\newblock {\em arXiv preprint arXiv:2006.14015}, 2020.

\bibitem{rasmussen2010gaussian}
Carl~Edward Rasmussen and Hannes Nickisch.
\newblock {Gaussian processes for machine learning (GPML) toolbox}.
\newblock {\em Journal of machine learning research}, 11(Nov):3011--3015, 2010.

\bibitem{rebentrost2022quantum}
Patrick Rebentrost, Alessandro Luongo, Samuel Bosch, and Seth Lloyd.
\newblock Quantum computational finance: martingale asset pricing for incomplete markets.
\newblock {\em arXiv preprint arXiv:2209.08867}, 2022.

\bibitem{alphatron}
Patrick Rebentrost, Miklos Santha, and Siyi Yang.
\newblock Quantum alphatron.
\newblock {\em Quantum 7 (2023): 1174.}, 2023.

\bibitem{reynolds2009gaussian}
Douglas~A Reynolds.
\newblock Gaussian mixture models.
\newblock {\em Encyclopedia of biometrics}, 741(659-663), 2009.

\bibitem{roosta2015improved}
Farbod Roosta-Khorasani and Uri Ascher.
\newblock Improved bounds on sample size for implicit matrix trace estimators.
\newblock {\em Foundations of Computational Mathematics}, 15(5):1187--1212, 2015.

\bibitem{rue2005gaussian}
Havard Rue and Leonhard Held.
\newblock {\em Gaussian Markov random fields: theory and applications}.
\newblock CRC press, Cambridge, 2005.

\bibitem{sachdeva2013faster}
Sushant Sachdeva and Nisheeth~K Vishnoi.
\newblock Faster algorithms via approximation theory.
\newblock {\em Theoretical Computer Science}, 9(2):125--210, 2013.

\bibitem{saibaba2017randomized}
Arvind~K Saibaba, Alen Alexanderian, and Ilse~CF Ipsen.
\newblock Randomized matrix-free trace and log-determinant estimators.
\newblock {\em Numerische Mathematik}, 137(2):353--395, 2017.

\bibitem{shepherd2006computation}
Dan Shepherd.
\newblock Computation with unitaries and one pure qubit.
\newblock {\em arXiv preprint quant-ph/0608132}, 2006.

\bibitem{shor2007estimating}
Peter~W Shor and Stephen~P Jordan.
\newblock Estimating jones polynomials is a complete problem for one clean qubit.
\newblock {\em Quantum Information \& Computation}, 8(8):681--714, 2008.

\bibitem{spielman2008graph}
Daniel~A Spielman and Nikhil Srivastava.
\newblock Graph sparsification by effective resistances.
\newblock In {\em Proceedings of the fortieth annual ACM symposium on Theory of computing}, pages 563--568, 2008.

\bibitem{stathopoulos2013hierarchical}
Andreas Stathopoulos, Jesse Laeuchli, and Kostas Orginos.
\newblock Hierarchical probing for estimating the trace of the matrix inverse on toroidal lattices.
\newblock {\em SIAM Journal on Scientific Computing}, 35(5):S299--S322, 2013.

\bibitem{gilbertstrang}
Gilbert Strang.
\newblock {\em Introduction to linear algebra}.
\newblock Wellesley-Cambridge Press, Wellesley MA, 2023.

\bibitem{subramanian2019quantum}
Sathyawageeswar Subramanian and Min-Hsiu Hsieh.
\newblock Quantum algorithm for estimating $\alpha$-renyi entropies of quantum states.
\newblock {\em Physical Review A}, 104(2):022428, 2021.

\bibitem{sun2021asymptotically}
Xiaoming Sun, Guojing Tian, Shuai Yang, Pei Yuan, and Shengyu Zhang.
\newblock Asymptotically optimal circuit depth for quantum state preparation and general unitary synthesis.
\newblock {\em arXiv preprint arXiv:2108.06150}, 2021.

\bibitem{sun2021querying}
Xiaoming Sun, David~P Woodruff, Guang Yang, and Jialin Zhang.
\newblock Querying a matrix through matrix-vector products.
\newblock {\em ACM Transactions on Algorithms (TALG)}, 17(4):1--19, 2021.

\bibitem{suri2011counting}
Siddharth Suri and Sergei Vassilvitskii.
\newblock Counting triangles and the curse of the last reducer.
\newblock In {\em Proceedings of the 20th international conference on World wide web}, pages 607--614, 2011.

\bibitem{suzuki2020amplitude}
Yohichi Suzuki, Shumpei Uno, Rudy Raymond, Tomoki Tanaka, Tamiya Onodera, and Naoki Yamamoto.
\newblock Amplitude estimation without phase estimation.
\newblock {\em Quantum Information Processing}, 19:1--17, 2020.

\bibitem{tang2024cs}
Ewin Tang and Kevin Tian.
\newblock A cs guide to the quantum singular value transformation.
\newblock In {\em 2024 Symposium on Simplicity in Algorithms (SOSA)}, pages 121--143. SIAM, 2024.

\bibitem{ubaru2017fast}
Shashanka Ubaru, Jie Chen, and Yousef Saad.
\newblock {Fast Estimation of tr$(f(A))$ via Stochastic Lanczos Quadrature}.
\newblock {\em SIAM Journal on Matrix Analysis and Applications}, 38(4):1075--1099, 2017.

\bibitem{udell2019big}
Madeleine Udell and Alex Townsend.
\newblock Why are big data matrices approximately low rank?
\newblock {\em SIAM Journal on Mathematics of Data Science}, 1(1):144--160, 2019.

\bibitem{van2020quantum}
Joran Van~Apeldoorn, Andr{\'a}s Gily{\'e}n, Sander Gribling, and Ronald de~Wolf.
\newblock Quantum sdp-solvers: Better upper and lower bounds.
\newblock {\em Quantum}, 4:230, 2020.

\bibitem{vos2016methods}
Vaya Sapobi~Samui Vos.
\newblock {\em Methods for determining the effective resistance}.
\newblock PhD thesis, Masters thesis, 20 December, 2016.

\bibitem{wang2014clustering}
Guan-Wei Wang, Chun-Xia Zhang, and Jian Zhuang.
\newblock Clustering with prim’s sequential representation of minimum spanning tree.
\newblock {\em Applied Mathematics and Computation}, 247:521--534, 2014.

\bibitem{GuomingWang}
Guoming Wang.
\newblock Efficient quantum algorithms for analyzing large sparse electrical networks.
\newblock {\em arXiv preprint arXiv:1311.1851}, 2013.

\bibitem{wang2022new}
Qisheng Wang, Ji~Guan, Junyi Liu, Zhicheng Zhang, and Mingsheng Ying.
\newblock New quantum algorithms for computing quantum entropies and distances.
\newblock {\em arXiv preprint arXiv:2203.13522}, 2022.

\bibitem{wang2019witnessing}
W~Wang, J~Han, B~Yadin, Y~Ma, J~Ma, W~Cai, Y~Xu, L~Hu, H~Wang, YP~Song, et~al.
\newblock Witnessing quantum resource conversion within deterministic quantum computation using one pure superconducting qubit.
\newblock {\em Physical Review Letters}, 123(22):220501, 2019.

\bibitem{wu2016estimating}
Lingfei Wu, Jesse Laeuchli, Vassilis Kalantzis, Andreas Stathopoulos, and Efstratios Gallopoulos.
\newblock Estimating the trace of the matrix inverse by interpolating from the diagonal of an approximate inverse.
\newblock {\em Journal of Computational Physics}, 326:828--844, 2016.

\bibitem{xie2016weighted}
Yuan Xie, Shuhang Gu, Yan Liu, Wangmeng Zuo, Wensheng Zhang, and Lei Zhang.
\newblock Weighted schatten $ p $-norm minimization for image denoising and background subtraction.
\newblock {\em IEEE transactions on image processing}, 25(10):4842--4857, 2016.

\bibitem{xin2018nuclear}
Tao Xin, Bi-Xue Wang, Ke-Ren Li, Xiang-Yu Kong, Shi-Jie Wei, Tao Wang, Dong Ruan, and Gui-Lu Long.
\newblock Nuclear magnetic resonance for quantum computing: Techniques and recent achievements.
\newblock {\em Chinese Physics B}, 27(2):020308, 2018.

\bibitem{yoganathan2019one}
Mithuna Yoganathan and Chris Cade.
\newblock The one clean qubit model without entanglement is classically simulable.
\newblock {\em arXiv preprint arXiv:1907.08224}, 2019.

\bibitem{yosef2024multivariate}
Liron~Mor Yosef, Shashanka Ubaru, Lior Horesh, and Haim Avron.
\newblock Multivariate trace estimation using quantum state space linear algebra.
\newblock {\em arXiv preprint arXiv:2405.01098}, 2024.

\bibitem{zhang2011laplacian}
Xiao-Dong Zhang.
\newblock The laplacian eigenvalues of graphs: a survey.
\newblock {\em arXiv preprint arXiv:1111.2897}, 2011.

\bibitem{zhang}
Yunong Zhang and William~E Leithead.
\newblock Approximate implementation of the logarithm of the matrix determinant in gaussian process regression.
\newblock {\em Journal of Statistical Computation and Simulation}, 77(4):329--348, 2007.

\bibitem{zhao2019compiling}
Liming Zhao, Zhikuan Zhao, Patrick Rebentrost, and Joseph Fitzsimons.
\newblock Compiling basic linear algebra subroutines for quantum computers.
\newblock {\em Quantum Machine Intelligence}, 3(2):1--10, 2021.

\bibitem{zhao2019quantum}
Zhikuan Zhao, Jack~K Fitzsimons, Michael~A Osborne, Stephen~J Roberts, and Joseph~F Fitzsimons.
\newblock Quantum algorithms for training gaussian processes.
\newblock {\em Physical Review A}, 100(1):012304, 2019.

\end{thebibliography}

\appendix

\section{Further preliminaries}

\subsection{Model of computation - quantum access to classical data}\label{apx:qaccessclassical}

In this section, we succinctly explain the QRAM model, and the quantum arithmetic model, and show how to build block-encodings of matrices. The runtimes of the algorithms we propose are expressed in the number of usages of a unitary which is block-encoding a matrix. A block-encoding of a matrix can be built using queries to a QRAM oracle. Hence, the total gate complexity of the computation is given by the number of queries to the oracles multiplied by the complexity needed to implement a single query. Informally, the QRAM model is the following. 

\begin{definition}[Informal - QRAM model \cite{kerenidis2020quantum,allcock2023constant}]\label{qrammodeldefinition}
An algorithm in the QRAM data structure model that processes data of size $m$ has two steps: 
\begin{enumerate} 
\item A preprocessing step with complexity $\widetilde{O}(m)$ that constructs efficient QRAM data structures for storing the data. 
\item A computational step where the quantum algorithm has access to the QRAM data structures constructed in step $1$. 
\end{enumerate} 
The complexity of the algorithm in this model is measured by the cost for step $2$.
\end{definition}

For a more precise definition of a quantum computer with quantum access to classical memory (in short, the QRAM model), the interested reader is referred to~\cite{allcock2023constant}, where the authors extend the standard definition of a quantum computer with the addition of a new gate. This gate enacts a query to a $\mathsf{QMD}$, a quantum memory device, which implements a user-programmable mapping. In this framework, the QRAM is a particular kind of $\mathsf{QMD}$. In this model, the cost of step $2$ is measured by the depth of the QRAM circuit, which --- in many implementations --- is logarithmic in the size of the data $m$.  The interested reader is referred to~\cite{hann2021resilience,giovannetti2008architectures,jaques2023qram} for more information about QRAM circuits with logarithmic depth. In this model, the quantum computer is organized in an input register (to model a possible initial state of the computation which might come from a quantum network), a \emph{workspace} (where the computation happens), an \emph{address register} and a \emph{target register}, storing respectively the address of the memory element to retrieve and the register to write it. The address and the index register are shared between the quantum computer and the $\mathsf{QMD}$. In the registers of the quantum computer, we can apply any gate from a fixed universal gate set.  The $\mathsf{QMD}$ consists in a \emph{memory register} (whose size is exponential in the length of the address register) and an \emph{ancilla register} (whose size is proportional to the size of the memory). Importantly, we can apply multi-qubit gates between qubits of the index, target, ancilla, and memory register, but we cannot apply multi-qubit gates between the workspace and the memory or the ancilla register of the $\mathsf{QMD}$. \\

Now we discuss how to build block-encodings using a QRAM. First, we have to construct a data structure (often called KP-trees) and store it in the QRAM. For a matrix $A \in \mathbb{R}^ {n \times d}$ with  $\|A\|_0$ non-zero elements, this preprocessing takes a classical preprocessing time and space of $O(\|A\|_0\log(nd))$. Once this preprocessing is done, the data structure is stored in the QRAM as per Definition~\ref{qrammodeldefinition}. Then, the \emph{quantum access to a matrix} allows us to create a block-encoding of $A$. More formally.

\begin{claim}[Quantum access to a matrix \cite{gilyen2019quantum,kerenidis2017recommendation, chakraborty2018power}]\label{def:quantumaccess}
Let $A=\in \mathbb{R}^{n \times d}$ that has been preprocessed and stored in a QRAM. A quantum algorithm with quantum access to this data structure can create $(\alpha(A)$-block-encoding of a matrix $A$, where $\alpha(A)$ is defined in Definition~\ref{def:mu}. There exists a circuit with depth of $O(\text{poly}(\log(nd)))$ and size of $O(nd\log(nd))$. 
\end{claim}

\begin{definition}[Parameter $\alpha(X)$ \cite{kerenidis2020quantum}]
\label{def:mu}
Let $A \in \mathbb{R}^{n \times d}$. 
Define $\alpha(A) = \min_ {p \in [0,1]}\left(\norm{X}_F, \sqrt{s_{2p}(X)s_{2(1-p)}(X^T)}\right)$, with $s_{q}(X) = \max_{i} \norm{X_{i,\cdot}}_q^q$, i.e. the maximum $q$-th power of the $q$-norm of the rows of $X$.
\end{definition}

We emphasize that, as we do for classical algorithms in the runtimes of our algorithms we neglect the dependence in the digits of precision used to store the entries of the matrix $a_{ij}$. We can always pad the matrix with zeros if we work with non-square matrices. In conclusion, we work under the assumption that the circuit depth required to build a block-encoding is polylogarithmic in the size of the matrix.

There are other models for creating block-encodings of a matrix. For sparse matrices, we can create block-encodings using \cite[Lemma 48]{gilyen2019quantum}, a procedure implemented and detailed in \cite{camps2022explicit}. In the standard sparse access model to a matrix (also known as oracle access in adjacency list model) is it possible to create $(\sqrt{s_rs_c}, \epsilon)$-block-encoding of $A \in \mathbb{R}^{2^a \times 2^a}$ using $O(\log(1/\epsilon))$ single and two qubits gates, where $s_r,s_c$ is the maximum sparsity of the rows and the columns~\cite{gilyen2019quantum,chakraborty2018power}. Also in this model, it is simple to preprocess the unitary giving sparse access so we obtain a block-encoding of a matrix with spectral norm bounded by $1$.

Finally, we can create a block-encoding of a density matrix from a state preparation unitary that gives access to a purification of the density matrix. This model reduces the preparation of a block-encoding to the problem of state preparation, for which we have efficient circuits \cite{sun2021asymptotically}.

\begin{lem}[Block-encoding of density operators {\cite[Lemma 45]{gilyen2019quantum}}]\label{lemma:purification} 
Suppose that $\rho$ is an $s$-qubit density operator, $G$ is an $(a+s)$-qubit unitary that on 
 the $\ket{0}\ket{0}$ input state prepares a purification $\ket{0}\ket{0}\mapsto \ket{\rho}$, s.t. $Tr_a\ket{\rho}\bra{\rho}=\rho$. Then $(G^\dagger \otimes I_s)(I_a \otimes SWAP_s)(G\otimes I_s)$ is a $(1,a+s,0)$-block-encoding of $\rho$.
\end{lem}

\subsection{Model of computation - arithmetic model}\label{apx:arithmetic model}

In addition to the model of quantum computation with quantum access to classical data, we recall the definition of a potential quantum arithmetic model. An arithmetic model is essential for specifying how numbers are represented in a quantum computer. The following presents one such arithmetic model, which focuses on real numbers. Additionally, other arithmetic models can be defined for processing integers, such as the two's complement and one's complement representations.

\begin{definition}[Fixed-point encoding of real numbers~\cite{alphatron}] \label{defEncoding}
Let $c_1,c_2$ be positive integers, and $a\in\{0,1\}^{c_1}$, $b \in \{0,1\}^{c_2}$, and $s \in \{0,1\}$ be bit-strings. Define the rational number as:
\begin{equation}
    \mathcal{Q}(a,b,s):= 
    (-1)^s
    \left(2^{c_1-1}a_{c_1}+ \dots + 2a_2 + a_1 + \frac{1}{2}b_1 + \dots + \frac{1}{2^{c_2}}b_{c_2} \right) \in [-R,R],
\end{equation}
where $R := 2^{c_1}-2^{-c_2}$. If $c_1,c_2$ are clear from the context, we can use the shorthand notation for a number $z:=(a,b,s)$ and write  $\mathcal{Q}(z)$ instead of $\mathcal{Q}(a,b,s)$. Given an $n$-dimensional vector $v \in (\{0,1\}^{c_1} \times \{0,1\}^{c_2} \times \{0,1\})^n$
the notation $\mathcal{Q}(v)$ means an $n$-dimensional vector whose $j$-th component is $\mathcal{Q}(v_j)$, for $j \in[n]$. 
\end{definition}

This means that, when we have to store or represent numbers in the \emph{workspace} of our quantum computer, we decide the number of bits of precision we require. For our purposes, we use definition~\ref{defEncoding} when we perform the trace estimation of lemma~\ref{lemma:quantum trace estimator vanilla} and lemma~\ref{lem:quantum trace estimator relative}. For a constant $c_2$, the smallest positive number we can represent is $\mathbf{m}_0 = 2^{-c_2}$.

\subsection{Useful quantum subroutines}\label{appx:usefulquantum}
For practicality, we recall here the statement of amplitude estimation: 
\begin{thm}[Quantum amplitude estimation \cite{brassard2002quantum}]\label{thm:amplitude_estimation}
There is a quantum algorithm which takes as input one copy of a quantum state $|\psi\rangle$, a unitary transformation 
$U = 2|\psi\rangle\langle \psi| - I$, a unitary transformation $V = I - 2P$ for some projector $P$, and an integer $t$. The algorithm outputs $\widetilde{a}$, an estimate of $a = \langle \psi|P|\psi\rangle$, such that 
\be \label{Quantum amplitude estimation}
|a - \widetilde{a}| \leq \frac{2\pi \sqrt{a(1-a)}}{t} + \frac{\pi^2}{t^2}
\ee
with probability at least $8/\pi^2$, using $U$ and $V$ $t$ times each.
\end{thm}

\subsubsection*{Product of block-encodings}
 
In this paragraph, we state some results on how to combine of block-encodings. We start from the simplest case where we combine the unitaries corresponding to two different block-encodings - and this will lead to a unitary that is the block-encoding of the product. Then, we show how to perform block-encoding amplification, and how this can be useful when combining block-encodings. We generalized a known previous result on the product of $2$ preamplified block-encodings to the product of $k$ preamplified block-encodings.

\begin{thm}[Product of block-encodings {\cite[Lemma 15]{chakraborty2022quantum}, \cite{gilyen2019quantum}}] 
\label{thm:prod of block-encodings} 
If $U_A$ is an $(\alpha, a, \delta)$-block-encoding of an s-qubit operator $A$ implemented in time $T_A$, and $U_B$ is a $(\beta, b, \epsilon)$-block-encoding of an s-qubit operator $B$ implemented in time $T_B$, then $(I^{\otimes b}\otimes U_A)(I^{\otimes a}\otimes U_B)$ is an $(\alpha\beta, a+b, \alpha\epsilon + \beta\delta)$-block-encoding of $AB$ implemented at a cost of $O(T_A + T_B)$.
\end{thm}

\begin{thm}[Uniform Block Amplification {\cite[Corollary 8]{chakraborty2022quantum}}\label{thm:uniform block amplification 2022}]
Let $A^{N \times d}$ and $\delta \in (0,1]$. 
Suppose $U$ is a $(\alpha, a, \epsilon)$-block-encoding of $A$, such that $\epsilon \leq \frac{\delta}{2}$, that can be implemented at a cost of $T_U$. 
Then a $(\sqrt{2}\|A\|, a+1, \delta)$-block-encoding of $A$ can be implemented at a cost of $O(\frac{\alpha T_U}{\|A\|}\log(\frac{\|A\|}{\delta}))$.
\end{thm}

As described in \cite{chakraborty2022quantum} the following theorem can be generalized to the product of matrices of any size $a \times b$ and $b \times c$.

\begin{thm}[Product of two preamplified block-encodings {\cite[Lemma 16]{chakraborty2022quantum}}\label{thm:product preamplified be 2022}]
Let $\delta \in (0,1]$. If $U_A$ is an $(\alpha_A, a_A, \epsilon_A)$-block-encoding of an $s$-qubit operator $A$ implemented in time $T_A$, and $U_B$ is a $(\alpha_B, a_B, \epsilon_B)$-block-encoding of an $s$-qubit operator $B$ implemented in time $T_B$, such that $\epsilon_A \leq \frac{\delta}{4\sqrt{2}\|B\|}$, $\epsilon_B \leq \frac{\delta}{4\sqrt{2}\|A\|}$. Then we can implement a $(2\|A\|\|B\|, a_A+a_B+2, \delta)$-block-encoding of $AB$ implemented at a cost of 
$$O\left(\left(\frac{\alpha_A}{\|A\|}T_A + \frac{\alpha_B}{\|B\|}T_B \right) \log\left(\frac{\|A\|\|B\|}{\delta} \right) \right)$$
\end{thm}

\begin{thm}[Product of $k$ amplified block-encodings] 
\label{thm:prod of k amplified block-encodings} 
Let $\delta \in (0,1]$. If $U_i$ is an $(\alpha_i, a_i, \epsilon_i)$-block-encoding of an s-qubit operator $A_i$ implemented in time $T_i$.
If $\epsilon_i \leq \frac{\epsilon}{2k(\sqrt{2})^{k-1}(\prod_{j\in[k], j\neq i} \norm{A_j})}$, then we can implement a $(2^{k/2}\prod_{i\in[k]}\|A_i\|, \sum_{i \in [k]}a_i + k, \epsilon)$-block-encoding of $\prod_{i \in [k]}A_i$ at a cost of $$O\left(\left(\sum_{i \in [k]}\frac{\alpha_i}{\norm{A_i}}T_i\right)\log\left(\frac{k(\sqrt{2})^{k-1}\prod_{i\in [k]}\|A_i\|}{\epsilon}\right)\right)$$.
\end{thm}
\begin{proof}
We first amplify each block-encoding using Corollary \ref{thm:uniform block amplification 2022} (uniform block amplification), and then combine them using theorem~\ref{thm:prod of block-encodings} (product of block-encodings).
    Amplifying each $U_i$ has a cost of $O\left(\frac{\alpha_i}{\norm{A_i}}T_i\log \left(\frac{\norm{A_i}}{\delta_i}\right)\right)$ respectively and gives us $(\sqrt{2}\norm{A_i}, a_{i}+1, \delta_{i})$-block-encodings for $\epsilon_i\leq \delta_i/2$
    Combining all the preamplified block-encoding gives a $(2^{k/2}\prod_{i=1}^k \norm{A_i}, k+\sum_{i=1}^k a_i, (\sqrt{2})^{k-1}\sum_i^k(\prod_{j\in[k], j\neq i} \norm{A_j})\delta_i)$-block-encoding of the product.
    We pick $\delta_i \leq \frac{\epsilon}{k(\sqrt{2})^{k-1}(\prod_{j\in[k], j\neq i} \norm{A_j})}$,  which bounds the final block-encoding error by $\epsilon$.
\end{proof}

\subsubsection*{Positive powers of block-encodings.}  The difference between the following two lemmas is the following. In the first, the matrix needs to be Hermitian and normalized such that $\|A\| \leq 1$. The procedure gives a $1$-block-encoding of $H^c/2$. The second algorithm, besides a logarithmic improvement in the runtime, gives a $\alpha^c$-block-encoding of $H^c/2$.

\begin{thm}[Positive powers of Hermitian matrices {\cite[Lemma 10]{chakraborty2018power}}]
\label{lemma: BE of matrix power hermitian}
Let $c \in (0, 1], \kappa \geq 2$, and
$H$ be a Hermitian matrix such that $I/\kappa \preceq H \preceq I$. Suppose that $\delta = o(\epsilon/(\kappa \log^3(\kappa/\epsilon)))$, and
we are given an $(\alpha,a,\delta)$-block-encoding of $H$ that is implemented in $O(T)$ elementary gates. Then 
for any $\epsilon$, we can implement a unitary that is a $(1, a + O(\log\log(1/\epsilon)), \epsilon)$-block-encoding of $H^c/2$ in cost
$$O(\alpha \kappa (a+T) \log^2(\kappa/\epsilon)).$$
\end{thm}

\begin{thm}
[Positive powers of any matrix {\cite[Corollary 30]{chakraborty2022quantum}}]
\label{lemma: be of any matrix power}
Let $c \in (0,1]$, be a constant. Let $A$ be a matrix with condition number $\kappa$. Let $U_A$ be a $(\alpha, a, \epsilon)$-block-encoding of $A$ implemented in time $T_A$, such that $\epsilon \leq \frac{\gamma\|A\|^{1-c}}{2c\kappa^{1-c}}$. Then we can construct a $(2\alpha^c, a+1, \delta)$-block-encoding of $A^c$ at cost 
$$O\left( \frac{\alpha\kappa}{\|A\|}\log\left(\frac{\alpha}{\delta}\right)T_A \right).$$
\end{thm}

\subsection{Polynomial approximations and useful claims}\label{appx:polyapprox}

In the following, we report some polynomial approximations of useful functions that are used in the main text. Remark that the original polynomial approximation of the logarithm function presented in \cite{distributional} (lemma~\ref{lemma:poly approx ln distributional} here) is required to be symmetric. We do not need this requirement, but we need for the polynomial to be bounded in absolute value by $1/2$ on the interval $[-1,1]$. This is also changing the error we can tolerate.

\begin{lem}[Polynomial approximation of $1/x$ {\cite{childs2017quantum}}]
\label{lemma:polynomial of inverse}
Let $\epsilon,\delta\in(0,1/2]$. Then there is a polynomial $\widetilde{V}$ of degree $O(\frac{1}{\delta}\log(\frac{1}{\delta\epsilon}))$ such that $ |\widetilde{V}(x) - {3\delta}/{8x} |\leq \epsilon$ on the domain  $[-1,1] \backslash [-\delta,\delta]$, moreover $|\widetilde{V}(x)|\leq 1/2$ for all $ x\in [-1,1]$.
\end{lem}
The previous lemma can potentially be improved in our setting using the methodology of \cite{gribling2021improving}.

\begin{lem}[Series of $-\ln(1-x)$]\label{lemma:serieslog}
The complex series is defined by:
\begin{equation}
    \sum_{n=1}^{+\infty}\frac{x^n}{n}
\end{equation}
has radius of convergence $1$ and converges to the function $-\ln(1-x)$ for all $-1\leq x<1$. Hence we obtain
that $-\ln(1+x) = \sum_{n=1}^{+\infty}(-1)^n\frac{x^n}{n}$ for all $-1<x\leq 1$.
\end{lem}

\begin{lem}[Polynomial approximation of logarithm\cite{distributional}]\label{lemma:poly approx ln distributional}
Let $\beta\in(0,1]$, $\epsilon\in(0,{1}/{6}]$. Then there exists a polynomial $\tilde{S}$
of degree $O({\frac{1}{\beta}\log (\frac{1}{\epsilon} )})$ such that $|\tilde{S}(x)-\frac{\log_b(x)}{3\log_b(2/\beta)}|\leq\epsilon$ for all $ x\in [\beta,1]$ and base $b \in \mathbb{N}$, and for all $ x\in [-1,1]$ $1/2
\leq \tilde{S}(x) = \tilde{S}(-x) \leq 1/2$.
\end{lem}

\begin{proof}
Recall lemma~\ref{lemma:serieslog}. 
We follow the same steps of the proof of \cite[Lemma 11]{distributional}. 
We consider the standard Taylor expansion of $\frac{\log(x)}{3\log(2/\beta)}$. centered in $x_0=1$, which is $f(x)=\frac{1}{3 \log(2/\beta)}\sum_{n \geq 1} \frac{(-1)^{1+n}(-1+x)^n}{n}$. We use this polynomial in \cite[Corollary 16]{distributional} with the choice of $\epsilon=\eta/2$, $x_0=1$, $r=1-\beta$, $\nu = \frac{\beta}{2}$. This corollary gives us another polynomial $S \in \mathbb{C}[x]$ of degree $O(\frac{1}{\beta}\log(\frac{1}{\eta}))$ with the following properties: 
\begin{align}
    \|f(x) - S(s)\|_{[\beta, 2-\beta]} \leq \eta/2 
    \end{align}
    \begin{align}\label{eq:fkB}
    \|S(x)\|_{[-1, 1]} \leq B+\eta/2 \leq \frac{1}{3}+\frac{\eta}{2} 
    \end{align}
    \begin{align}
    \|S(x)\|_{[-1, \beta/2]} \leq \eta/2
\end{align}
Again, following the steps of the original proof~\cite[Corollary 66]{gilyen2019quantum}, we now show that indeed $B=1/3$ in~\ref{eq:fkB}, and thus the polynomial is bounded by $1/2$ on the $[-1,1]$ interval. We have to consider $f(x_0 +x) = \frac{1}{3 \log(2/\beta)}\sum_{n=0}^{\infty} = \frac{a_n (-1+x+x_0)^n}{n}$, and find a bound for $\sum_{n=0}^\infty (r+\nu)^n|a_n|$, which in our case (as $a_n = (-1)^{1+n}/n$) is $\frac{1}{3\log(2/\beta)}\sum_{n=0}^\infty \frac{(1-\beta/2)^n}{n}=1/3$.

\end{proof}

\begin{lem}[Polynomial approximation of monomials on {$[-1, 1]$} \cite{sachdeva2013faster}]\label{lemma:poly approx of monomial} 
For positive integers $s,d$, there exist an efficiently computable polynomial $E_{s,d}(x) \in \mathbb{R}[x]$ of degree $d$ such that:
\[
\sup_{x\in[-1,1]} |E_{s,d} (x) - x^s| \leq 2e^{{-d^2}/{2s}}.
\]
\end{lem}

\begin{thm}[Folklore]\label{thm:usefulbound}
$(a^p-b^p)=(a-b)(\sum_{i=0}^{p-1}a^{i}b^{p-1-i} )$
\end{thm}

\begin{theorem}[\cite{rebentrost2022quantum}]\label{lem:martingaleobservation}
Let $A \in \mathbb{R}^{n \times d}$ with singular value decomposition $A= UDV^\dagger$.
The condition number of $A$ as $\kappa(A) = \|A\|\|A^+ \|$.  Let $S \subseteq [d]$ a subset of the indices of the columns and $A_S$ the matrix obtained by picking only the columns of $A$ with index in $S$. Then $\kappa(A_S) \leq \kappa(A)$.    
\end{theorem}

\begin{claim}
\label{claim:exponential-error}
Let $a,\overline{a} > 0$ such that $|a - \overline{a}|\leq \epsilon/2$. Then $|e^a - e^{\overline{a}}| \leq e^a \epsilon$.
\end{claim}
\begin{proof}
The proof follows from the mean value theorem, which  in our case states that for a continuous function over  $[a-\epsilon, a+\epsilon]$ and differentiable over $(a-\epsilon, a+\epsilon)$, there exists a $c$ in the open set such that 
$$\frac{d}{dx}e^c = \frac{e^{a} - e^{\overline{a}}}{a - \overline{a}}.$$
Thus, $e^c (a - \overline{a}) = e^{a} - e^{\overline{a}}$. Taking the absolute value on both sides, we obtain that $ |e^{a} - e^{\overline{a}}| \leq e^{a+\epsilon/2}|a-\overline{a}| \leq e^{a+\epsilon/2}\epsilon/2 \leq e^a(1+\epsilon/2)\epsilon/2 = e^a\epsilon/2 + e^a\epsilon^2/4 \leq e^a\epsilon$ . 

\end{proof}

\section{Trace estimation subroutines}\label{apx:trace estimation subroutines}
In this section, we restate in the language of block-encodings some well-known results for estimating the trace of a matrix $A$, see for example ref.~\cite{shor2007estimating}. This is stated in lemma~\ref{lemma:quantum trace estimator vanilla}. Using some standard trick in computer science --- which we borrow from \cite{subramanian2019quantum, chowdhury2021computing}, albeit with some minor modifications and --- we obtain the estimate of the trace with relative error, which we formalize in lemma~\ref{lem:quantum trace estimator relative}. We report in theorem~\ref{corollary: trace estimation of product final} the statement of the theorem for estimating the trace of $A^TA$, which was known from~\cite{van2020quantum}. For these algorithms, the generalization to matrices in $\mathbb{C}$ is straightforward.

Interestingly, there are also classical algorithms that can obtain an estimate in absolute error with only $O(1/\epsilon)$ quantum queries to an oracle, similarly to the quantum algorithm~\cite{meyer2021hutch++}.

\begin{lem}[Quantum trace estimation (absolute error)]
\label{lemma:quantum trace estimator vanilla}
Let $\epsilon \in (0,1)$. 
Let $U$ be a $(\alpha, q, \delta)$ block-encoding of $A \in \mathbb{C}^{n \times n}$. There is a quantum algorithm that returns $\overline{{\rm Tr}[A]}$ with  probability at least $2/3$ such that $\left|{\rm Tr}[A] - \overline{{\rm Tr}[A]}\right| \leq  n \epsilon$  using $O\left(\frac{\alpha}{\epsilon}\right)$ calls of $U$ and $U^\dagger$, if $\delta \leq \epsilon/2$.
The probability can be made bigger than $1-\varepsilon$ for $\varepsilon \in (0, 1/2)$ with a multiplicative factor of $O\left(\log(1/\varepsilon) \right)$.
\end{lem}

\begin{proof}
The algorithm consists in performing amplitude estimation on the circuit of the Hadamard test. 
In the top-left corner of $U$ we have $\widetilde{A}/\alpha$. Denote $\ket{\widetilde{\psi}}$ the state obtained by applying $U$ to the first two registers of $\ket{\phi} = ({1}/{\sqrt{n}})\sum_{i=0}^{n-1} \ket{i}\ket{0^{\otimes q}}\ket{i}$ (where the second register is corresponds to the $q$ additional qubits of the block-encoding), i.e. 
$\ket{\widetilde{\psi}} = \ket{0}\frac{\widetilde{A}\otimes I}{\alpha}\ket{i}\ket{0^{\otimes q}}\ket{i} + \ket{G,0^\bot}$, where $G$ is a non-normalized garbage state orthogonal to the first register.
Then, similarly to the Hadamard test, consider the state 
\be
\ket{\varphi} = \frac{1}{2} \ket{0} (\ket{\psi} + \ket{\phi})
+\frac{1}{2} \ket{1} (\ket{\psi} - \ket{\phi}).
\ee
This state is obtained by starting from the $\ket{0}\ket{0}$ state (the first register being a single qubit, and the second register having the same size of $\ket{\psi}$), and performing a Hadamard gate on the first register, and then, controlled on the first register being $0$ we create state $\ket{\psi}$, controlled on the qubit being $1$ create the state $\ket{\phi}$, and then we perform another Hadamard in the first qubit. 
Observe that 
\be 
\braket{\psi|\phi}=\frac{1}{n} \sum_{ij} \bra{j}\bra{j}\frac{\widetilde{A}}{\alpha}\ket{i}\ket{i} = \frac{1}{n} \sum_i \braket{i|\frac{\widetilde{A}}{\alpha}|i}=\frac{\rm Tr[\widetilde{A}]}{\alpha n}
\ee
and the probability of measuring $\ket{0}$ on the ancilla qubit equals $a := (1+\text{Re}(\Tr[\widetilde{A}])/\alpha n) / 2$.  Using amplitude estimation (theorem~\ref{thm:amplitude_estimation}), we can obtain $\widetilde{a}$ such that $|a-\widetilde{a}| \leq \epsilon/4\alpha$  by choosing $t = \frac{16\alpha\pi}{\epsilon}$. This gives an error $|a - \widetilde{a} |  \leq \epsilon/4\alpha$. Now set $\overline{{\rm Tr}[A]} = n\alpha(2\widetilde{a} - 1)$, then using triangle inequality,
\begin{align}
\Big| \overline{{\rm Tr}[A]} - \Tr[A] \Big| 
\leq &
\Big|\Tr[A] - \Tr[\widetilde{A}] \Big| + \Big| \overline{{\rm Tr}[A]} - \Tr[\widetilde{A}] \Big|  \nonumber \\
\leq &  n \delta + 2\alpha n|a -\widetilde{a}| \leq \frac{n\epsilon}{2} + \frac{n\epsilon}{2} \leq n \epsilon.
\end{align}

In the above chain of inequalities, we used $\delta \leq \epsilon/2\alpha$ from our hypotheses. The complexity of the above procedure is given by the runtime of performing amplitude estimation, which calls $U$ and $U^\dagger$ for $O(\alpha/\epsilon\sqrt{a})$ times, and we note that $a$ has a lower bound of $1/2$ when the trace is $0$. The failure probability can be pushed to be bigger than $1-\varepsilon$ by repeating the previous subroutine (which succeeds with constant probability greater than $8/\pi^2$) for $O(\log(1/\varepsilon)$ times, and taking the median of the estimates (i.e. this is a standard technique that goes under the name of powering lemma \cite{powering}).
\end{proof}

The following theorem is (a small generalization) from~\cite{van2020quantum} and \cite[Lemma 11.9]{thesissander}.

\begin{thm}[Trace estimation of product of matrices]
\label{corollary: trace estimation of product final}
Let $U$ be a $(\alpha,a,\delta)$-block-encoding of a matrix $B \in \C^{n\times m}$, with $n\leq m$, implementable in time $O(T_U)$.
Let $\epsilon \in \R^+$ and $\alpha, n$ %
be known.  
A multiplicative $\epsilon$-approximation of $\Tr[B^\dagger B]$ can be computed at a cost of $\widetilde{O} \left( \frac{\alpha\sqrt{n}}{\epsilon\sqrt{\Tr[B^\dagger B]}}T_U \right)$, assuming $\delta \leq \frac{\epsilon\sqrt{Tr[B^\dagger B]}}{4\sqrt{n}}$.
\end{thm}

\subsection{Trace estimation with relative error}
To obtain a relative error estimate on a quantity $a \in (0,1]$, using an algorithm that returns an absolute error estimates, we need a lower bound $\lambda$ on $a$. Then, we can set the estimation error to $\epsilon \lambda$, and we can simply observe that $|a- \overline{a}| \leq \epsilon \lambda \leq \epsilon a$. However, often we do not have a lower bound on $a$. In these cases, the solution consists in using a simple exponential search (a kind of binary search) to find it. Intuitively, the algorithm works by performing a search for a lower bound $\lambda$ for the value of $a$, and then we proceed by running $\text{ESTIMATE}()$ with error $\epsilon \lambda$. This is the idea of theorem~\ref{thm:fromabstorel} from \cite{chowdhury2021computing}, which we report and formalize here. If a crude lower bound is known, the analysis of the algorithm can be greatly simplified, and we discuss it in the last paragraph of this section. This is a very mild assumption to satisfy: in any real implementation of an algorithm, we have a lower bound given by the machine precision $\mathfrak{m}_0$. This intuition is clearly formalized by using an \emph{arithmetic model}, which is detailed in appendix~\ref{apx:arithmetic model}. In the following algorithm, it is possible that the function $\text{ESTIMATE}()$ might not use an upper bound on the quantity $a_{max}$.

\begin{thm}[Relative estimates from absolute estimates]\label{thm:fromabstorel}
    Let $\text{ESTIMATE}(\epsilon_{abs} > 0, \delta' > 0, a_{max} > 0) \mapsto \mathbb{R}$ be a function that returns an estimate $\overline{a}$ of a quantity $a \in (0, \infty)$ so that
    $$Pr[|a - \overline{a}| \leq \epsilon_{abs}] \geq 1-\delta'. $$
    Then, for $\epsilon \in (0,1], \delta \in (0, 1]$, and a known $ a_{max} > a$, algorithm~\ref{alg:abs2rel} estimates the quantity $a$ with relative error, i.e. such that 
        $$Pr[|a - \overline{a}| \leq \epsilon a ] \geq 1-\delta. $$

On the $r$-th iteration of algorithm~\ref{alg:abs2rel}  we call $\text{ESTIMATE}(\frac{\epsilon a_{max}}{2^r}, \frac{6\delta }{\pi^2r^2}, a_r)$, for an expected number of $O(\log_2(a_{max}/a))$ iterations. 
\end{thm}

\begin{algorithm}
\caption{Return relative error estimate from absolute error estimates}\label{alg:abs2rel}
\begin{algorithmic}
\Require $\text{ESTIMATE}(\epsilon'>0, \delta'>0, a_{max}>0)$, $\epsilon \in (0,1), \delta \in (0, 3/4]$, $a_{max} \in \mathbb{R}$ 
\Ensure $\overline{a} \in (0,  a_{max})$
\State $r=0$
\State $a_0 = a_{max}$
\State $\overline{a}_0 = 0$
\While{$a_r > \overline{a}_r$}
    \State $r = r+1$
    \State $a_r = a_{max}/2^r$, 
    \State $\epsilon_{abs}(r) = \epsilon_{rel}a_r/2$  
    \State $\delta'(r) = \frac{6}{\pi^2}\frac{\delta}{r^2}$
    \State $\overline{a}_r = \text{ESTIMATE}(\epsilon_{abs}(r), \delta'(r), a_{max})$ 
\EndWhile
\State \textbf{Return} $\overline{a}_r$.
\end{algorithmic}
\end{algorithm}

\begin{proof}
The proof is divided into two parts. First, we prove the correctness of the algorithm, and then we prove the runtime complexity of our statement. In the following, let $a_{r}$ the upper bound for $a$ at iteration $r$ and let $\overline{a}_r$ our estimate of $a$ at iteration $r$. We denote with $R$ the time when Algorithm~\ref{alg:abs2rel} stops, producing an estimate $\overline{a}_R$. We model the value $R$ and the value $\overline{a}_r$ as random variables. For ease of notation, we define $\overline{a}_R = \overline{a}$. 

The probability of the estimate $\overline{a}_R$ being correct is lower bounded by the probability that every sequence $\overline{a}_i$ is correct. Without any assumption on $R$ we can lower bound this quantity by 
    \begin{equation}
        \prod_{i=0}^\infty (1- \delta'(i)) \geq 1 - \delta \frac{6}{\pi^2}\sum_{i=0}^\infty \frac{1}{r^2}= 1-\delta 
    \end{equation}

The first inequality follows by induction on $i$, and in the equality we used the Euler series $\sum_{i=0}^\infty \frac{1}{i^2} = \frac{\pi^2}{6}$. The algorithm's output is such that
        $$Pr[|a - \overline{a}_R| \leq \epsilon_{abs}(R) ] \geq 1-\delta $$ 

Since $\epsilon_{abs}(R)= \epsilon_{rel}a_R/2$, and $\overline{a} \geq a_R$, it follows trivially that $|a - \overline{a} | \leq \epsilon_{rel}\overline{a}/2$, which equivalently can be written as 

\begin{equation}\label{eq:bound1}
\overline{a}(1-\frac{\epsilon_{rel}}{2}) \leq a \leq \overline{a}(1+\frac{\epsilon_{rel}}{2}).
\end{equation}

However, we want to obtain a bound in function of $a$, i.e. we want to find a small $c$ such that $|a-\overline{a}|\leq c a$, i.e. $\overline{a} \leq a(1+c)$ and $\overline{a} \geq a(1-c)$. Using simple arithmetic on eq.~\ref{eq:bound1} we can obtain 
\begin{equation}\label{eq:inverted-bound}
a\left(1-\frac{\epsilon_{rel}}{2}\right)^{-1} \geq \overline{a} \geq a\left(1+\frac{\epsilon_{rel}}{2}\right)^{-1}.
\end{equation}
We combine two of the previous inequalities to obtain

\begin{align}
(1+c) = & \left(1-\frac{\epsilon_{rel}}{2}\right)^{-1}  = \left(\sum_{j=0}^\infty \left(\frac{\epsilon}{2}\right)^j\right) = \left(1+\sum_{j=1}^\infty \left(\frac{\epsilon}{2}\right)^j\right) \\ 
\leq &  \left(1+\epsilon\sum_{j=1}^\infty \left(\frac{1}{2}\right)^j\right)\leq (1+\epsilon),
    \end{align}

where we used the geometric series in the second equality. Thus, a small $c$ such that $|a - \overline{a}| \leq ca$ is $\epsilon$. By doing the same bound on the other side of Equation~\ref{eq:inverted-bound}, we obtain that $|a-\overline{a}| \leq \epsilon a$. In conclusion, we have that 
        $$Pr[|a - \overline{a}_R| \leq \epsilon_{rel}a ] \geq 1-\delta.$$ 

This proves the correctness of Algorithm~\ref{alg:abs2rel}.  Now we can prove the runtime complexity of the algorithm, in terms of the number of queries to the $\text{ESTIMATE}()$ function. Remember that instead of computing the expected number of iterations $\mathbb{E}[R]=\sum_{R=1}^\infty Pr(R)R$, we just want to compute a bound for it. We proceed as follows. 
First observe that there exists an integer $q \geq 1$ such that
\begin{equation}\label{bounds-on-a}
\frac{a_{max}}{2^{q}} \leq a < \frac{a_{max}}{2^{q-1}}.
\end{equation}
Equivalently, $q-1 < \log(\frac{a_{max}}{a}) \leq q$. It follows that if we assume that $\text{ESTIMATE}()$ works without any error, we will stop at $q$ (if the error is small and $a_{max}/2^q \ll a$) or from $q+1$ onwards. This because, in the worst case, i.e. when $a_{max}/2^q = a$, using the assumption that $\epsilon_{rel}\leq 1$ and observing that $\min \widetilde{a}_r = a-a_{r-1}$, we always satisfy the stopping condition:
$$a_{q+1} = a_{q}- a_{q+1} = a-a_{q+1} \leq \min \widetilde{a}_q \leq \widetilde{a}_{q+1}.$$

To bound the expected runtime of our algorithm we recall that for a discrete and non-negative random variable, we can express $\mathbb{E}[X] =\sum_{q'=0}^\infty Pr[X>q']  = \sum_{q'=1}^\infty Pr[X \geq q']$. In our case, we want to bound $\mathbb{E}[R]$ by bounding the different $Pr[R \geq q']$: the probability that the algorithm stops at iteration $q'$ or after. We consider two cases.

\begin{itemize}
\item The probability of stopping at iteration $q'$ before the stopping condition is satisfied (i.e. at iteration $q$ or $q+1$) is $1-\delta'(q')$ (it can only stop if $\text{ESTIMATE}()$ fails). We can bound each of the $1-\delta(q')$ by $1$.
\item For iteration $q' \geq q+1$ the algorithm is expected to stop with high probability, and it won't stop only if $\text{ESTIMATE}()$ fails, which happens with probability below $\delta'(q')=\frac{\pi^2 \delta}{6(q')^2}$.
\end{itemize}

\begin{align}
  \mathbb{E}[R] =&  \sum_{q'=1}^\infty Pr[R \geq q'] \\
  = & \sum_{q'=1}^{q+1} Pr[R \geq q'] + \sum_{q'=q+2}^{\infty} Pr[R \geq q'] \nonumber \\
   \leq &  (q+1) + \sum_{q'=q+2}^{\infty} Pr[R \geq q'] 
   \end{align}

Now we bound the second term. Note that $Pr[R \geq q']$ is equivalent to the probability of \emph{not} stopping strictly before $q'$. As the random variable modeling the stopping probability at every iteration is independent of the stopping probability at another iteration, this probability is given by $\prod_{r=1}^{q'} Pr[R=r]$. As before, remember that we bound $\prod_{r=1}^{q+1} \delta'(r) \leq 1$, and therefore,

\begin{align}
    Pr[R \geq q'] <&  \prod_{r=1}^{q'-1} \delta'(r) < \prod_{r=q+1}^{q'-1} \delta'(r) =  \prod_{r=q+1}^{q'-1} \frac{6 \delta}{\pi^2 r^2}  \nonumber \\
    = &\left( \frac{6 \delta}{\pi^2} \right)^{q'-q-1} \prod_{r=q+1}^{q'-1} \left( \frac{1}{r} \right)^2  = \left( \frac{6 \delta}{\pi^2 } \right)^{q'-q-1} \left(  \frac{q!}{(q'-1)!} \right)^2  \nonumber \\
    \leq & \left( \frac{6 \delta}{\pi^2} \right)^{q'-q-1} \left(  \frac{1}{(q'-q -1)!} \right)^2 <  \left( \frac{6 \delta}{\pi^2} \right)^{q'-q-1} \left(  \frac{1}{(q'-q -1)!} \right) 
\end{align}
For the last inequalities, we observed that $\frac{a!}{b!} \leq \frac{1}{(b-a)!}$ for $b>a$ follows from the positivity of the binomial coefficient, and $p^2 < p$ for $p<1$. In conclusion, we can derive the bound as follows:
   \begin{align}
 \mathbb{E}[R] \leq &  (q+1) + \sum_{q'=q+2}^{\infty} \left( \frac{6 \delta}{\pi^2} \right)^{q'-q-1} \left(  \frac{1}{(q'-q -1)!} \right) \nonumber  \\
 = & (q+1) + \sum_{i=1}^\infty \left(\frac{6 \delta}{\pi^2} \right)^i \left(  \frac{1}{i!} \right) \nonumber \\
 \leq &  (q+1) + e^{6\delta/\pi^2} - 1 \leq q+2. 
\end{align}

Where in the last step we used the Taylor expansion of the exponential and the fact that $\delta \leq 1$.  Finally, the way we can bound the expected run time as:
\begin{equation}
    \mathbb{E}[R] \leq q+2 \leq \log_2(a_{max}/a) + 2.
\end{equation}
\end{proof}

In the following, we obtain a simple proof using theorem~\ref{thm:fromabstorel} (relative estimates from absolute estimates) inside the proof of lemma~\ref{lemma:quantum trace estimator vanilla} (quantum trace estimation with absolute error).

\begin{lem}[Quantum trace estimation with relative error]
\label{lem:quantum trace estimator relative} Let $\epsilon, \delta \in (0,1)$, and let $U_A$ be a $(\alpha, q, \epsilon_1)$ block-encoding of a matrix $A\in \mathbb{R}^{n \times n}$ such that $\Tr[A] \neq 0$ and $\|A\|\leq 1$,   
 and $\epsilon_1 \leq \frac{\varepsilon \Tr[A]\alpha}{2n}$. There is a quantum algorithm that returns $\overline{{\rm Tr}[A]}$ such that $\left|{\rm Tr}[A] - \overline{{\rm Tr}[A]}\right| \leq \epsilon \Tr[A] $ and uses $U$ and $U^\dagger$ for $O(\frac{\alpha n}{\Tr[A]\epsilon} \log(\frac{n\alpha}{{\Tr}[A]})\log(\frac{\log(\frac{n\alpha}{Tr[A]})}{\delta}) )$
\end{lem}

\begin{proof}
In the proof, we consider matrices with positive traces. The case for a negative or complex trace follows similarly. From the assumptions, we have $0< \Tr[A] \leq n=:a_{\max}$. In the top-left corner of $U_A$ we have $\widetilde{A}/\alpha$, and from the assumption that $\|A - \widetilde{A}\| \leq \epsilon_1$ it follows that $|\Tr[A] - \Tr[\widetilde{A}]| = |Tr[A-\widetilde{A}]| \leq \sum_i |\lambda_i -\widetilde{\lambda}_i| \leq \frac{\epsilon \Tr[A]\alpha}{2}$.

The quantum algorithm follows the same steps as the algorithm of theorem~\ref{lemma:quantum trace estimator vanilla}, but we use theorem~\ref{thm:fromabstorel}. In the proof of lemma~\ref{lemma:quantum trace estimator vanilla}, we estimate (using amplitude estimation - theorem~\ref{thm:amplitude_estimation})  $a := (1+\text{Re}(\Tr[\widetilde{A}])/\alpha n) / 2 \in (1/2, 1]$, from which we derive $\Tr[\widetilde{A}]/\alpha n = 2a-1 \in (0, 1]$.
The procedure that estimates  $\frac{\Tr[\widetilde{A}]}{\alpha n}$ is the output of our $\text{ESTIMATE}()$ function, which we use in theorem~\ref{thm:fromabstorel} (relative error for absolute estimates) to obtain a relative estimate $\epsilon/2$ with failure probability $\delta$, and $a_{max} = 1$. 
In the last iteration $R=O(\log_2(a_{max}/a))$ (in expectation) of theorem~\ref{thm:fromabstorel} we run $\text{ESTIMATE}()$ with error 
 $\epsilon_{abs} = \frac{\epsilon a_{max}}{4 \cdot 2^{R}} = \frac{\epsilon\Tr[A]}{4 n\alpha}$. Similarly to the previous proof, it follows from the triangle inequality that
\begin{align}
\Big| \overline{\frac{\Tr[\widetilde{A}]}{\alpha n}} - \frac{\Tr[A]}{\alpha n} \Big| 
\leq &
\Big|\frac{\Tr[A]}{n\alpha} - \frac{\Tr[\widetilde{A}]}{\alpha n} \Big| + \Big| \frac{\overline{{\rm Tr}[\widetilde{A}]}}{\alpha n} - \frac{\Tr[\widetilde{A}]}{\alpha n} \Big|  \nonumber \\
\leq &  \epsilon\frac{\Tr[A]}{2\alpha n} + \epsilon\frac{\Tr[A]}{2\alpha n} \leq  \epsilon \frac{\Tr[A]}{\alpha n}.
\end{align}

Observe that $\text{ESTIMATE}()$ will run amplitude estimation on the probability $a$ with error $\epsilon_{abs}/2$ (as $\Big| \frac{\overline{{\rm Tr}[\widetilde{A}]}}{\alpha n} - \frac{\Tr[\widetilde{A}]}{\alpha n} \Big| \leq 2|a-\overline{a}|$) which will result in 
 using $t=\lceil \frac{16 \pi \alpha n }{\epsilon \Tr[A]} \rceil$ in theorem~\ref{thm:amplitude_estimation}.
The total number of usages of $U_A$ and its inverse is bounded, in expectation, by

$$O\left( \frac{\alpha n}{\epsilon \Tr[A]} \log\left(\frac{n\alpha}{\Tr[A]}\right)\log\left(\frac{\log(\frac{n\alpha}{Tr[A]})}{\delta}\right)\right).$$

\end{proof}

\begin{remark}\label{remark:boundsonkappa}
In the context of theorem~\ref{lem:quantum trace estimator relative} (trace estimation with relative error), the runtime is bounded by $\widetilde{O}(\frac{\alpha \kappa}{\epsilon})$.
\end{remark}
\begin{proof}
The runtime for the relative error estimate can be bounded by $O(\frac{\kappa\alpha}{\epsilon})$, as it is simple to check that $\frac{n}{\Tr[A]} \leq \frac{n}{n \lambda_{min}} \leq \frac{\lambda_{max}}{\lambda_{min}} = \kappa$
\end{proof}

\subsection{From absolute to relative error in a quantum arithmetic model}
We can simplify algorithm~\ref{alg:abs2rel} if we work in a quantum arithmetic model, which we described in~\ref{apx:arithmetic model}. In particular, we can leverage the fact that there is inherently a smallest number that we can represent on a quantum (or classical) computer in binary encoding, and we call such number $\mathsf{m}_0$. This value represent a ``lower bound'' on the lower bound on the quantity we are estimating. Instead of changing the failure probability of the subroutine that returns an estimate with absolute error, we can keep this number fixed. Hence, we use $\mathsf{m}_0$ to define $R = \lceil \log \left(\frac{x_{max}}{\mathsf{m}_0} \right)\rceil$, and then use $R$ to set the probability of success in the algorithm~\ref{alg:abs2rel} to be $c':=1-\frac{(1-c)}{R} \geq \frac{3}{4}$ for a a $c\in (0,1)$. In this setting, the analysis of the algorithm is greatly simplified. If $\mathsf{m}_0$ is proportional to $a$, the dependency of the failure probability on the runtime of the algorithm can be improved compared to theorem~\ref{thm:fromabstorel}.

\section{Numerical experiments}\label{apx:numerical schatten}

We numerically measure the scaling of the factor $\rho=\frac{(\sqrt{2}\|A\|)^{p/2}}{\|A\|_p^{p/2}}$ for real-world matrices. For this experiment we consider\footnote{\url{https://github.com/Scinawa/SpectralSums}} matrices obtained from standard datasets in machine learning, and computed the value $\rho$. The value of $\rho$ is bounded above by the value of $\sqrt{2}^{p/2}$ and converges to $\sqrt{2}^{p/2}$, as the Schatten $p$-norm of a matrix $A$ converges to $\|A\|$ for $p \mapsto \infty$. We considered normalized matrices (i.e. $\|A\|=1$) obtained from various datasets, with some focus in anomaly detection. 
\begin{itemize}
    \item MNIST: a dataset of handwritten digits~\cite{lecun1998mnist}, a standard benchmark dataset in machine learning, used in pattern recognition. 
    \item Banknote: a dataset of genuine and forged banknote-like samples~\cite{lohweg2012banknote},
    \item Sparse: a dataset obtained by generating $10000$ samples of $50$ features using density of $10\%$ (i.e. $10\%$ of the entries are non-zero numbers between $0$ and $1$).
    \item Cardio: a healthcare dataset of $1831$ samples, $21$ features, with $9.61\%$ of anomalies~\cite{han2022adbench,ayres2000sisporto}.
    \item Fraud: a banking dataset of fraudulent credit card transaction
detection composed by $284807$ samples, $29$ features, and $0.17\%$ of anomalies~\cite{han2022adbench,pang2019deep}.
\end{itemize}

\begin{figure}[H]
\centering
\includegraphics[width=10cm]{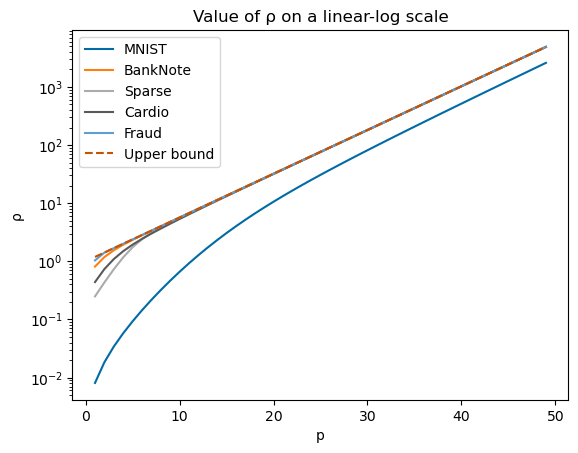}
\caption{Results of the numerical experiment measuring $\rho$ as function of $p \in [50]$.}
\label{fig:unicafigura}
\end{figure}

\end{document}